\documentclass[11pt,preprint]{aastex}
\usepackage{times}
\usepackage{graphicx}
\usepackage{amsmath}
\usepackage{multirow} 

\def\eg{{\it e.g.}}
\def\ltsima{$\; \buildrel < \over \sim \;$}
\def\simlt{\lower.5ex\hbox{\ltsima}}
\def\gtsima{$\; \buildrel > \over \sim \;$}
\def\simgt{\lower.5ex\hbox{\gtsima}}
\def\hide#1{}

\begin{document}

\def\eq{equation}
\def\HI{\mbox{\ion{H}{1}}}
\def\HII{\mbox{\ion{H}{2}}}
\def\GI{\mbox{\ion{He}{1}}}
\def\GII{\mbox{\ion{He}{2}}}
\def\GIII{\mbox{\ion{He}{3}}}
\def\etal{{\it et al.~\/}}
\def\cf{{\it cf.}}
\def\ie{{\it i.e.}}
\def\eg{{\it e.g.}}
\def\ltsima{$\; \buildrel < \over \sim \;$}
\def\simlt{\lower.5ex\hbox{\ltsima}}
\def\gtsima{$\; \buildrel > \over \sim \;$}
\def\simgt{\lower.5ex\hbox{\gtsima}}

\title{Where are the Fossils of the First Galaxies? II. True Fossils,
  Ghost Halos, and the Missing Bright Satellites} 
\author{Mia S. Bovill and Massimo Ricotti} \affil{Department of Astronomy,
  University of Maryland, College Park, MD 20740}
\email{msbovill@astro.umd.edu}

\begin{abstract}
  We use a new set of cold dark matter simulations of the local
  universe to investigate the distribution of fossils of primordial
  dwarf galaxies within, and around the Milky Way. Throughout, we
  build upon previous results showing agreement between the observed
  stellar properties of a subset of the ultra-faint dwarfs and our
  simulated fossils. Here, we show that fossils of the first galaxies
  have galactocentric distributions and cumulative luminosity
  functions consistent with observations. In our model, we predict  $\sim 300$ luminous satellites orbiting the Milky Way, $~50-70\%$ of
  which are well preserved fossils. Within the Milky Way virial
  radius, the majority of these fossils have luminosities
  $L_V<10^5$~L$_\odot$. Despite our multidimensional agreement with observations at low masses and luminosities the primordial model produces an overabundance of bright
  dwarf satellites ($L_V > 10^4 L_\odot$) with respect to observations where observations are
  nearly complete. The ``bright satellite problem'' is most evident in
  the outer parts of the Milky Way. We estimate that, although
  relatively bright, the primordial stellar populations are very
  diffuse, producing a population with surface brightnesses below
  surveys detection limits and are easily stripped by tidal
  forces. Although we cannot yet present unmistakable evidence for the
  existence of the fossils of first galaxies in the Local Group, the
  results of our studies suggest observational strategies that may
  demonstrate their existence. i) The detection of ``ghost halos'' of
  primordial stars around isolated dwarfs would prove that stars
  formed in minihalos ($M<10^8 M_\odot$) before reionization, and
  strongly suggest that at least a fraction of the ultra-faint dwarfs
  are fossils of the first galaxies. ii) The existence of a yet
  unknown population of $\sim150$ Milky Way ultra-faints with
  half-light radii $r_{hl}\approx100-1000$~pc and luminosities
  $L_V<10^4$~L$_\odot$, detectable by future deep surveys. These undetected dwarfs would have the mass-to-light ratios, stellar velocity dispersions and metallicities predicted in this work.
\end{abstract}

\section{Introduction}

Over the last decade, since \cite{Klypinetal:99} and
\cite{Mooreetal:99} showed that the number of dark matter subhalos
expected around a Milky Way mass halo is two orders of magnitude above
the number of known satellites, significant effort has been made in
observation and theory to solve the substructure problem in cold dark
matter (CDM) cosmology.  Observational discoveries have redefined the
'Missing Galactic Satellite Problem' with the number of observed Milky
Way satellites now somewhat closer to theoretical expectations
\citep{SimonGeha:07, Tollerudetal:08, BovillRicotti:09,
  Maccioetal:10}. The discovery of the ultra-faint dwarfs
\citep{Belokurovetal:06a, Belokurovetal:07, Irwinetal:07, Walshetal:07,
  Willmanetal:05AJ, Willmanetal:05ApJ,Zuckeretal:06b, Zuckeretal:06a,
  Gehaetal:09} has roughly doubled the known Milky Way and M31
satellite populations since 2004. However, observations alone cannot
fully explain the discrepancy between luminous satellites and CDM
substructure.

One of the core theoretical issues of the ``missing satellites''
problem remains the relationship between luminosity of satellites
and the virial mass at formation of their dark matter halos. From a
theoretical perspective, the fundamental question to be answered is:
``What is the minimum halo mass which can host a luminous galaxy?''
Previous studies \citep{Efstathiou:92, ThoulWeinburg:96, Bullocketal:01,
  Venkatesanetal:01, RicottiOstriker:04, RicottiOstrikerGnedin:05}
have shown that this critical mass is set by the reheating history of
the intergalactic medium (IGM) \citep{RicottiGnedinShull:00} and by
feedback loops operating before reionization which determine whether
low mass minihalos with $M<10^8 M_\odot$ are able to accrete gas from
the IGM and form stars. Simulations show that for halos with masses $M
< 10^8$~M$_\odot$ the local and stochastic components of galaxy
feedback produce minihalos with the same dark matter mass, but with
stellar masses which vary by several orders of magnitude
\citep{RicottiGnedinShull:08}. It is likely unjustified to assume a
sharp mass threshold separating dark and luminous halos and/or
a tight relationship between dwarfs' luminosities and their total mass,
at the faint end of the luminosity function. This is one of the main
motivations for the present study.

Cooling in halos with masses at formation, $M > 10^8$~M$_\odot$
($v_{max}>20$~km/s) is initiated via readily available hydrogen
Lyman-$\alpha$ emission. However, cooling in minihalos with $T_{vir} <
10^4$~K requires the formation of either molecular hydrogen or
pre-enrichment with metals from nearby galaxies. The balance between
the destruction and formation of $H_2$ and metal transport in the IGM
governs whether or not low mass minihalos can initiate cooling and form
stars. If ultraviolet $H_2$ dissociating radiation dominates, star
formation in the first minihalos may be suppressed or delayed
\citep{Haimanetal:00,Ciardietal:00,Machaceketal:00, WiseA:07, OSheaN:08,
  Maioetal:10}. However, if ionizing ultraviolet radiation ($h\nu >
13.6$~eV) is the dominant feedback mechanism, $H_2$ formation can be
catalyzed inside relic HII regions and on the edges of Stromegen
spheres \citep{RicottiGnedinShull:01, RicottiGnedinShull:02a,
  RicottiGnedinShull:02b, Ahnetal:06, Whalenetal:08, WiseA:08},
allowing star formation to be more widespread in minihalos with mass
$M \simlt10^8$~M$_\odot$, before reionization. It is important to
emphasize that, regardless of the interplay between the creation and
destruction of $H_2$, not every minihalo will host a galaxy.

The ultra-faint dwarfs are an excellent laboratory for testing models
of star formation in the lowest mass halos, due to their extremely low
luminosities: $10^2 - 10^5$~L$_\odot$, low dynamical masses: $M \simlt
10^7$~M$_\odot$ \citep{Strigarietal:08,Walkeretal:10,Wolfetal:10}, and their
estimated large number \citep{Koposovetal:07, Tollerudetal:08,
  Walshetal:09}. In \cite{BovillRicotti:09} and
Bovill \& Ricotti (2010a) (hereafter Paper I), we argue that the
properties of the ultra-faint dwarfs are consistent with those of
simulated primordial fossil galaxies.  \cite{Kravtsov:10} has argued
that the missing galactic satellites can be accounted for by a simple
model in which the fraction of baryons turned into stars, $f_\ast$
decreases with the virial mass in a halo before its infall into a
Milky Way. Not surprisingly, our simulations of pre-reinization dwarf
galaxies naturally produce such a decrease of star formation
efficiency with mass, although with increasing scatter at small masses
\citep{RicottiGnedinShull:08}.

Assuming the observed ultra-faints belong to an isotropically
distributed satellite population around the Milky Way \citep[but
see,][]{Metzetal:07,Metzetal:09,Bailinetal:08}, one infers from
observations a minimum of 60-65 satellites within $200$~kpc of the
Galactic center. This number is consistent with all the satellites
being subhalos that formed after reionization. There are about 90 (60)
subhalos that during their evolution had a maximum circular velocity
above $20$~km/s ($30$~km/s), thus likely formed most of their stars
after reionization \citep{BovillRicotti:09} (hereafter BR09). However,
we know that most satellites with $L_V<10^4$~L$_\odot$ are
undetectable beyond $50-100$~kpc \citep{SimonGeha:07}. Once luminosity
corrections are folded into estimates of the satellite counts, the
number of satellites inferred from observations is $\sim200-250$
\citep{Tollerudetal:08,Walshetal:09}, suggesting that around the Milky
Way there could be $\sim100-150$ true pre-reionization fossils
\citep[for a review see,][]{Ricotti:10}. Nevertheless, these
luminosity corrections are uncertain and model dependent, hence it is
largely unknown how many ultra-faints may exist in the outer parts of
the Milky Way. Indeed, the dimmest ultra-faint dwarfs within $50$~kpc
of the Milky Way may belong to a different population: according to
our simulations, they have properties that are inconsistent with well
preserved fossils, suggesting that may have been shaped
by tidal forces. Some observations also seem to reinforce this
hypothesis \citep{McGaughW:10, Willmanetal:10, Sandetal:09,
  Frebeletal:10}. In summary, although there is circumstantial
evidence that a fraction of the new ultra-faint dwarfs represent a
well preserved primordial population in halos with infall masses corresponding to $v_{max}\simlt20-30$~km~s$^{-2}$, we have yet to find observational proof of
their existence.

But is this ``primordial scenario'' consistent with observations when
we move beyond the virial radius of our own galaxy and peer into the
voids?  Unless the local component of galaxy feedback is very strong,
star formation should proceed similarly in small mass halos regardless
of a halo's location relative to the Local Group. Therefore, the voids
should be populated with luminous objects (see Figure 1 of Paper I).
As first noted in \cite{Peebles:01}, they are not.  The number of
dwarf galaxies with absolute magnitude $M_V > -16$ ($L_V < 2\times10^8
L_\odot$) observed in the voids is smaller than expected in CDM
cosmologies \citep{Karachentsevetal:04, Karachentsevetal:06,
  Tullyetal:06}. According to \cite{TikhonovKlypin:09}, the luminosity
function can only be reconciled if halos with $v_{max} <
35$~km~s$^{-1}$ are dark.  However, such a large mass threshold for
star formation would produce less than 35 luminous satellites between
100-200 kpc of the Milky Way. That is inconsistent with observations
unless satellites with $L_V< 10^4$~L$_\odot$ do not exist beyond
100~kpc (see Figs.~3-4 in this paper).

Alternatively, if the star formation rate is primarily determined by
halo mass, the void phenomenon can be reconciled with CDM by using a
halo occupation distribution for which the $M/L$ ratio increases with
decreasing halo mass.  However, this solution has only been tested for
halos with $M > 10^{10}$~M$_\odot$ \citep{TinkerConroy:09}, three
orders of magnitude more massive than our fossils. When we extend
their $M/L \sim M^{-1}$ relation to our mass range we obtain a $M/L$
ratio $\simgt 10^5$ for the fossil population ($M \simlt 10^8
M_\odot$). This would produce an ``ultra-faint'' population with $M/L$
ratios $\sim2-3$ orders of magnitude greater than those seen for the
primordial fossils in \cite{RicottiGnedinShull:02b}

In this paper, we investigate the following conundrum: can we
simultaneously account for the predicted and observed subhalo
population around the Milky Way and the lack of isolated galaxies with
$M_V > -16$ in the voids? Is a primordial origin scenario for ultra-faint
dwarfs, consistent with the number of dwarfs in the voids? The simulations
described in Paper I allow us to address both of these populations
within the same theoretical framework. As a result of this study, we
will propose three new observational strategies to search for evidence
of fossil galaxies in the local universe. We will argue that, if stars
formed in halos with masses $M<10^8 M_\odot$ there should be diffuse
``ghost halos'' of primordial stars around isolated dwarfs, in
addition to the undetected population of lower surface brightness
ultra-faint dwarfs.

This paper is structured as follows. \S~\ref{Numerical}
and~\ref{Obser} summarize the numerical method and treatment of the
observations, covered in detail in Paper~I. In \S~\ref{RES} we discuss
the radial distribution of the fossils of the first galaxies and make
comparisons to other N-body simulations (\S~\ref{RDs}) and we show
agreement between the radial distributions and cumulative luminosity
functions of primordial fossils in our simulations and the observed
ultra-faint satellites (\S~\ref{LFs}). We also discuss a possible
origin of Willman~I type ultra-faint dwarfs and the exception of Segue
I. In \S~\ref{LuSP}, we describe an overabundance of bright satellites
in the outer parts of the Milky Way halo and investigate possible
solutions. In \S~\ref{Tests}, we propose observational tests for our
models. Discussion and a summary are presented
in \S~\ref{Disc}.

\section{Numerical Method}
\label{Numerical}

In this paper, we continue our analysis of the simulations described
in Paper I. In brief, we used a set of hybrid initial conditions for a
$\Lambda$CDM N-body simulation to give us the resolution necessary to
study the primordial fossils on Local Volume scales.  Our high
resolution region, centered on a `Local Group,' is generated as
follows. Instead of using a uniform grid of particles, on scales
$l<1$~Mpc, the positions and velocities of our halos are set by the
final outputs of the \cite{RicottiGnedinShull:02a,
  RicottiGnedinShull:02b, RicottiGnedinShull:08} simulations,
hereafter referred to as the pre-reionization simulations.
Perturbations on scales $l>1$~Mpc are added to the initial positions
and velocities of particles at $z=z_{init}$ using a coarse resolution
simulation run from $z=40$ to $z=0$. For more detail, we direct the
reader to the appendix of Paper I. In addition to the non-uniform
initial grid of particles, the particles in the high resolution region
have masses set by the mass function of the final pre-reionization
outputs. Thus, we do not need to resolve these pre-reionization halos,
just trace their merger history, tidal disruption, and positions from
reionization to present day.

Paper I details the two realizations of our initial conditions Our
first order realization (see Appendix in Paper I for details) uses the
same pre-reionization output across the entire high resolution region,
and consequently does not account for the slower rate of structure
evolution in the voids. In contrast, our second order initial
conditions use four outputs from the pre-reionization simulations and
do account for slower rate of structure formation in the voids
\citep{BarkanaL:04}. Near the Milky Way, where this paper is
concerned, we find no substantive difference between the two
realizations. We therefore focus our discussion on the second order
initial conditions.

All the simulations discussed in this paper were run using Gadget 2
\citep{Springel:05} and analyzed using the halo finder AHF
\citep{KnollmannKnebe:09}. Simulation and data analysis were run on
the beowulf cluster Deepthought at the HPCC at the University of
Maryland. At $z=0$ we define a bound halo as anything identified by
AHF that, for the parameters we used, produces mass functions complete
for halos with a number of particles $N>50$. Paper I shows that our
hybrid initial conditions and non-uniform particle masses produce results
consistent with traditional CDM N-body simulations on both Local
Volume and galactic subhalo scales.  The parameters of the three
`Milky Ways' contained within our two highest resolution runs are
listed in Table~\ref{MWtab}.

\begin{deluxetable}{ccccc}
\tablecaption{Table of 'Milky Ways'}\label{MWs}
\tablecolumns{7}
\tablewidth{0pt}
\tablehead{
\colhead{Name} &
\colhead{Run} &
\colhead{Mass} &
\colhead{$R_{vir}$} &
\colhead{$v_{max}$} \\
\colhead{} &
\colhead{} &
\colhead{($10^{12} M_\odot$)} &
\colhead{(kpc)} &
\colhead{(km s$^{-1}$)} 
}
\startdata
MW.1 & C & 1.82 & 248 & 203 \\
MW.2 & D & 0.87 & 222 & 196 \\ 
MW.3 & D & 1.32 & 194 & 177 \\ 
\enddata
\label{MWtab}
\end{deluxetable}

\subsection{Fossil Definition}\label{Fossil}

Dark matter halos identified in our simulations at $z=0$ are divided
into three populations based on their ability to accrete gas from the
IGM and form stars after reionization. A just virialized halo is able
to accrete gas after reionization only if its maximum circular
velocity, $v_{max}$, is larger than a filtering velocity,
$v_{filt}$. In this work, we use $v_{filt}=20-30$~km~s$^{-1}$,
corresponding to the threshold for cooling via Lyman-$\alpha$. Paper I
and GK06 show that the exact choice of the filtering velocity does not
significantly change the results. If a $z=0$ halo has $v_{max} >
v_{filt}$, we classify it as a {\it{non-fossil}}. In the modern epoch,
non-fossils can be identified as dIrrs which have been accreting gas
and forming stars continuously since reionization.

Any halos whose present day $v_{max}$ is below the filtering velocity
is a {\it{candidate fossil}}. A candidate fossil for which
$v_{max}>v_{filt}$ at any point during its evolution may have accreted
gas and/or formed stars after reionization. We include them in the
``non-fossil'' group, as post-reionization stars may be the
dominant stellar population. When we need to distinguish these
``candidate fossils'' from the broader ``non-fossil'' group, we will
refer to them as {\it{polluted fossils}}.  Finally, any candidate
fossil for which $v_{max}<v_{filt}$ from reionization to the modern
epoch is a {\it{true fossil}}. True fossils formed the majority
($>70\%$) of their stars before reionization and today would be
relatively diffuse systems of old stars devoid of gas
\citep{RicottiGnedin:05}. The term {\it{fossil}} will only refer to
``true fossils''. Unless otherwise specified, the term non-fossil will
apply to any halo which could have accreted gas after reionization
regardless of its maximum circular velocity at $z=0$.

The division of the Milky Way and M31 satellites into fossils and non-fossils is shown in Table 2. We distinguish between the seven, classical RG05 fossils above the $10^6 L_\odot$ threshold, the RG05 fossils with $L_V<10^6 L_\odot$ and the ultra-faint dwarfs discovered since 2005.

\begin{table*}
\centering
\begin{tabular}{| c || c | c c c |}
\hline
& \multirow{3}{*}{Non-fossils} & \multicolumn{3}{| c |}{Fossils} \\
& & RG05 only & RG05 $\&$ BR11a & BR11a only \\
\hline
\hline
\multirow{10}{*}{Milky Way} & LMC & Sculptor & Draco$^{1}$ & Bootes I $\&$ II\\
& NGC 55 & & Phoenix & CVn I $\&$ II\\
& Sextans A $\&$ B & & Sextans & Hercules\\
& SMC & & Ursa Minor & Leo IV $\&$ T$^2$ \\
& WLM & & & Pisces II \\
& {\it{Carina}} & & &\\
& {\it{Fornax}} & & & \\
& {\it{GR8}} & & &\\
& {\it{Leo I, II $\&$ A}} & & &\\
& {\it{Sagittarius}} & & &\\
\hline
\multirow{11}{*}{M31} & IC 10 & And I $\&$ II & And V & And XI  XII\\
& IC 1613 & And III & & And XIII $\&$ IV \\
& IC 5152 & And VI & & And XV $\&$ XVI \\
& M32 & Antila & & AndXVII $\&$ XVIII \\
& NGC 185 & KKR 25 & & And XX $\&$ And XXI \\
& NGC 205 & & & And XXII $\&$ XXIII \\
& NGC 3109 & & & And XXIV $\&$ XXV \\
& NGC 6822 & & & And XXVI $\&$ XXVII \\
& {\it{DDO 210}} & & &\\
& {\it{LGC3}} & & &\\
& {\it{Pegasus}} & & &\\
\hline
\multirow{2}{*}{Isolated} & \multirow{3}{*}{---} & \multirow{3}{*}{---} & Cetus$^3$ & \multirow{3}{*}{---} \\
& & & Tucana &\\
\hline
\end{tabular}
\caption{Table of all the Local Group dwarfs divided by their host (or lack there of) and their non-fossil or fossil status. We have further divided the fossils into three groups, those considered fossils in RG05, but with $L_V>10^6 L_\odot$, the ``classical'' fossil dwarfs with luminosities below the $10^6 L_\odot$ threshold from BR11a, and, finally, those satellites which were only included in BR11a,b due to their post-2004 discovery. Three of dwarf initially identified as fossils show interesting properties (1) a small fraction of the stars in Draco are of intermediate age \citep{CioniHabing:05}, (2) Leo T has $\sim10^5 M_\odot$ of gas and a young stellar population \citep{deJongetal:08}, and (3) Cetus shows evidence for star formation through $z\sim1$ \citep{Monellietal:10}.}
\label{TAB.fos}
\end{table*}

\section{A Note on Observations}
\label{Obser}

We approach the observations as follows. The majority
of the information on the classical dwarfs comes from the
\cite{Mateo:98} review. For the ultra-faint dwarfs we generally defer
to measurements with the smallest error bars with some weight given to
more recent work (\cite{Walkeretal:09}). We direct
the reader to Paper I of this series and BR09 for a more complete
discussion of these criteria.

When calculating the observed distributions of dwarfs around the Milky
Way, we account for two effects, the sky coverage of the SDSS, and its
detection efficiencies \citep{Walshetal:09,Koposovetal:07}. For the
classical dwarfs, we assume the entire sky has been covered and only
apply sky coverage corrections to the ultra-faint population. To
correct for the SDSS sky coverage, we assume that the satellite
distribution around the Milky Way is isotropic, and multiply the
number of ultra-faints by 3.54 to account for the nearly
three-quarters of the sky not surveyed by SDSS, now past Data Release
7 \citep{Abazajianetal:09}. However, bright satellites of the Milky Way are distributed very anisotropically \citep{Kroupaetal:05, Zentneretal:05,Metzetal:07,Metzetal:09,Bozeketal:11AAS}, so the assumption of isotropy may not be a good one. 

Next, we apply a correction for the detection efficiency of the SDSS
using the results from \cite{Walshetal:09}. If an ultra-faint is
bright enough to be detected with $99\%$ efficiency, we assume the
sample is complete for that luminosity and distance. However, if the
ultra-faint is too dim for $99\%$ detection, but bright enough to be
detected half the time, we assume that, statistically, there is
another satellite with similar luminosity and distance missed by
SDSS. This second correction produces only a minor increase in the
number of satellites; approximately one additional satellite over a
total of $\sim60$ from sky coverage correction alone.

\begin{table*}
\centering
\begin{tabular}{| c || c | c |}
\hline
& $R < 50$~kpc & $R > 50$~kpc \\
\hline
\hline
\multirow{3}{*}{Inconsistent} & Segue 1 & \multirow{2}{*}{Pisces II} \\
& Segue 2 & \\
& Willman 1 & Leo V $^\ast$ \\
\hline
\multirow{5}{*}{Consistent} & \multirow{5}{*}{Coma Ber.} & Bootes I $\&$ II \\
& & CVn I $\&$ II \\
& & Hercules \\
& & Leo V $\&$ Leo T \\
& & Ursa Major I  $\&$ II \\
\hline
\end{tabular}
\caption{Table of Milky Way ultra-faint dwarfs classified by their distance from our galaxy (columns) and whether or not they are consistent with our predictions for the fossils of the first galaxies (row). Note the correlation between distance and consistency. (*) Pisces II and Leo V are both on the lower end of radii expected for fossils, as such they are marked as ``inconsistent,'' but are not as far from predictions as the ``inconsistent'' ultra-faints within $50$~kpc.}
\label{TAB.tidal}
\end{table*}

As discussed in Section 4 of Paper I, we also divide the ultra-faint dwarfs into two groups. The first is a group of seven, including CVnI and II, Hercules, Leo IV, Leo T, and Ursa Major I and II, have half light radii and surface brightnesses which are consistent with the stellar properties of fossils of the first galaxies. In contrast, the second group composed of five members, including Willman 1, Segue 1 and 2, Leo V, and Pisces II, have half light radii which are too small, and surface brightnesses which are too high, to be consistent with the simulated primordial population. 

Table~\ref{TAB.tidal} shows the Milky Way ultra-faint dwarfs classified in two ways, (i) the satellites within and beyond $50$~kpc, and (ii) those which are consistent (opened circles in right panel of Figure 11 of BR11a) and those which an inconsistent (filled green circles in the same figure) with the expected properties of fossil dwarfs. Excepting Leo V and Pisces II, the dwarfs which have half-light radii significantly smaller than the simulations are also within $50$~kpc of the Milky Way. Conversely, all the ultra-faint dwarfs which are consistent with the simulations are beyond $50$~kpc, except Coma Ber.. In this paper, we will refer to this second population collectively as the ``inner ultra-faints,'' to emphasize their location within the Local Group. 

Though we are agnostic about its status, Segue 1 may be an exception. Recent work \citep{Simonetal:10,Martinezetal:10} suggest that its stellar population has remained well within its tidal radius (thus tides are not important) and its stars are unaffected by interactions with the Milky Way. However, other work suggests that Segue 1 is a highly disrupted star cluster or dwarf \citep{Niederste-Ostholtetal:09,Norrisetal:10}. We note that, {\it{if}} Segue 1 is a undisrupted dwarf, the high concentration which has protected Segue I's stars also identifies it as a rare object formed in a high sigma peak at high redshift. The $1$~Mpc$^{3}$ volume of our pre-reionization simulations does not represent a large enough volume to contain a Segue 1. If Segue 1 is an undisrupted dwarf, than yes, if there are more than one or two additional Segue 1 like objects in the Milky Way halo it is a problem for our model that produces larger half-light radii than Segue 1's. However, if Segue 1 is disrupting than (i) we would not expect to see objects of that type beyond $~100$~kpc from the Milky Way, and (ii) the presence of additional Segue 1 objects would not pose a problem.s

The ultra-faint dwarfs in the first group, and the classical dSph noted in
\cite{RicottiGnedin:05} are the best candidates for an observed
population of primordial fossils, with stellar spheroids not
significantly modified by tides. However, as noted in Paper~I,
classical dwarfs with $L_V > 10^6 L_\odot$ are too bright to be hosted
in halos with $v_{max}<v_{filt}$ for $v_{filt}=20$~km~s$^{-1}$ or
$30$~km~s$^{-1}$. While they may have formed most of their stars
before reionization, we exclude them from out comparison to be as
conservative as possible.

Throughout this work, and in Paper I, we compare the observed Milky Way
satellites to our luminous $z=0$ halos with $\Sigma_V >
10^{-1.4}$~L$_\odot$~pc$^{-2}$. We are also able to use our
simulations to study the distribution of a hereto undetected
population of ultra-faints with $\Sigma_V <
10^{-1.4}$~L$_\odot$~pc$^{-2}$ and $L_V \simlt 10^4$~L$_\odot$.  The
possible existence and undetectability of this population was first
noticed in BR09, from the analysis of RG05 simulations (see also
Ricotti 2010 for a review). However, using independent arguments,
\cite{Bullocketal:10} have also proposed the existence of this
population they refer to as ``stealth galaxies.''

\section{Results}\label{RES}

In this section, we compare the distributions of non-fossils and
true-fossils to the galactocentric radial distribution of the observed
Milky Way satellites. We first compare the galactocentric radial
distributions of our simulations to observations.  We then make
detailed comparisons between the observed cumulative luminosity
function of the Milky Way satellites and the simulated cumulative
luminosity functions of our non-fossil and true fossil
populations. Note, that our simulated cumulative luminosity functions
only include stellar populations formed before
reionization. Therefore, we refer to our simulated cumulative
luminosity functions as primordial cumulative luminosity functions. Any star
formation that may take place in halos with $v_{max}>v_{filt}$ after
reionization is not accounted for in our simulated luminosity functions. Thus, only the cumulative luminosity function of
true fossils can be directly compared to observations, while the
luminosities of the non-fossils are lower limits.

\subsection{Radial Distribution of Fossils Near Milky Ways}
\label{RDs}

\begin{figure*}[t]
\plottwo{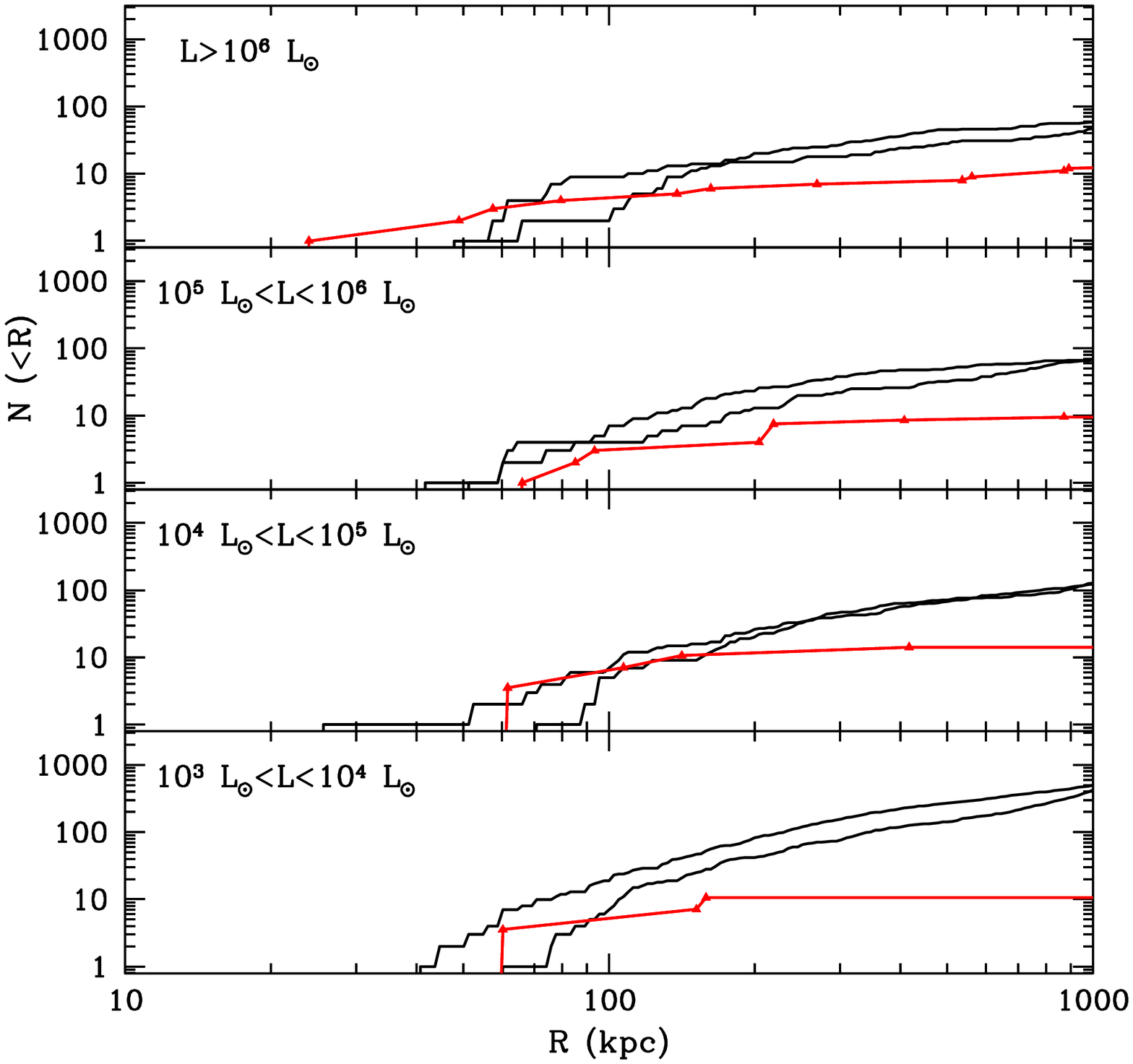}{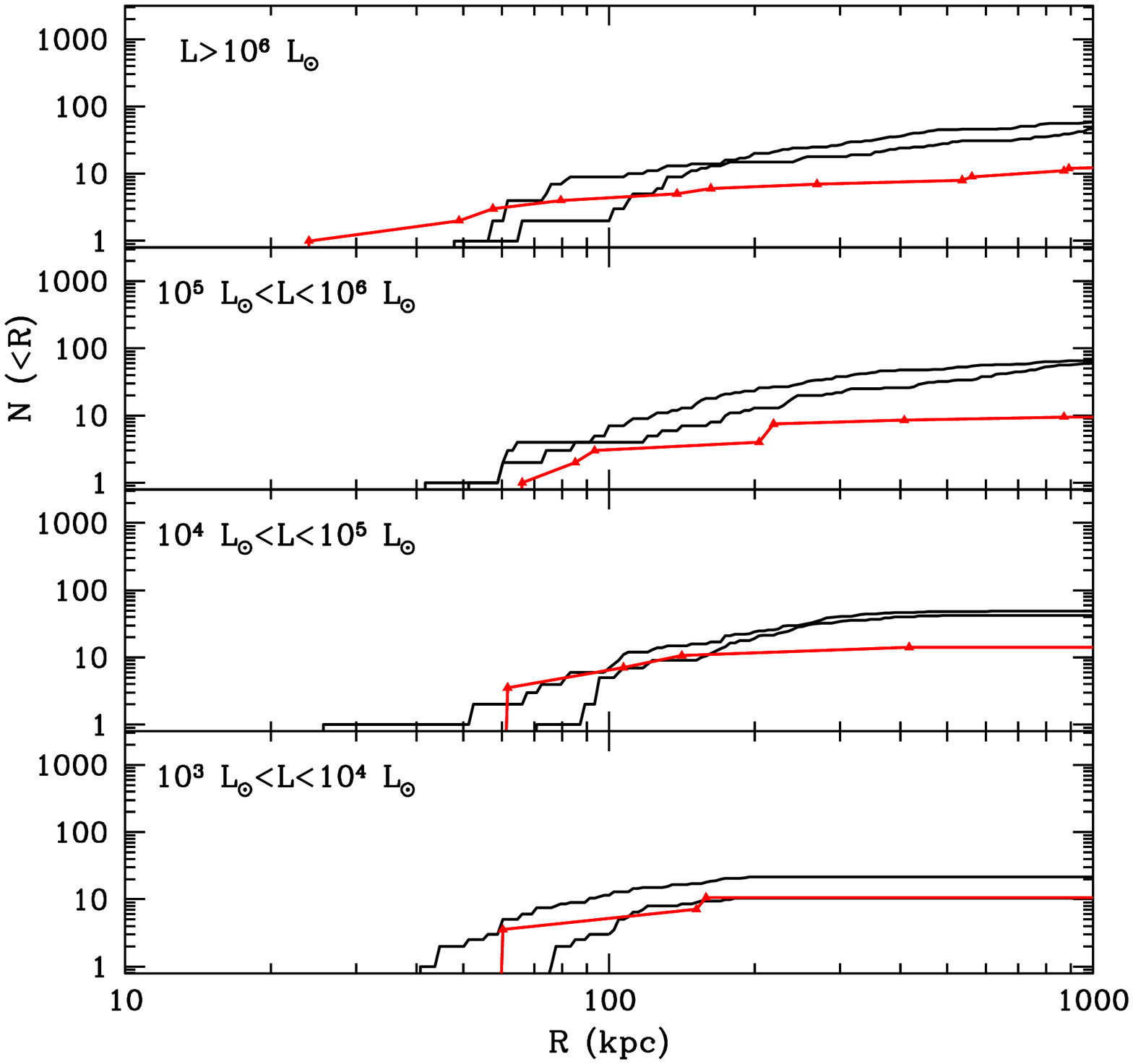}
\caption{{\it{Left}} Galactocentric radial distribution of all simulated satellites for
  MW.2 and MW.3 from Run D (black curves) compared to the radial
  distribution of all observed Milky Way satellites (red triangles).  We have
  included all simulated subhalos and known satellites regardless of
  their classification or whether they are detectable. {\it{Right}}
  Same as the left panel but we have convolved our populations with \cite{Walshetal:09} detection limits and only included simulated subhalos
  which can be detected in the SDSS data.}
\label{RD.all}
\end{figure*}

\begin{figure*}[t]
\plottwo{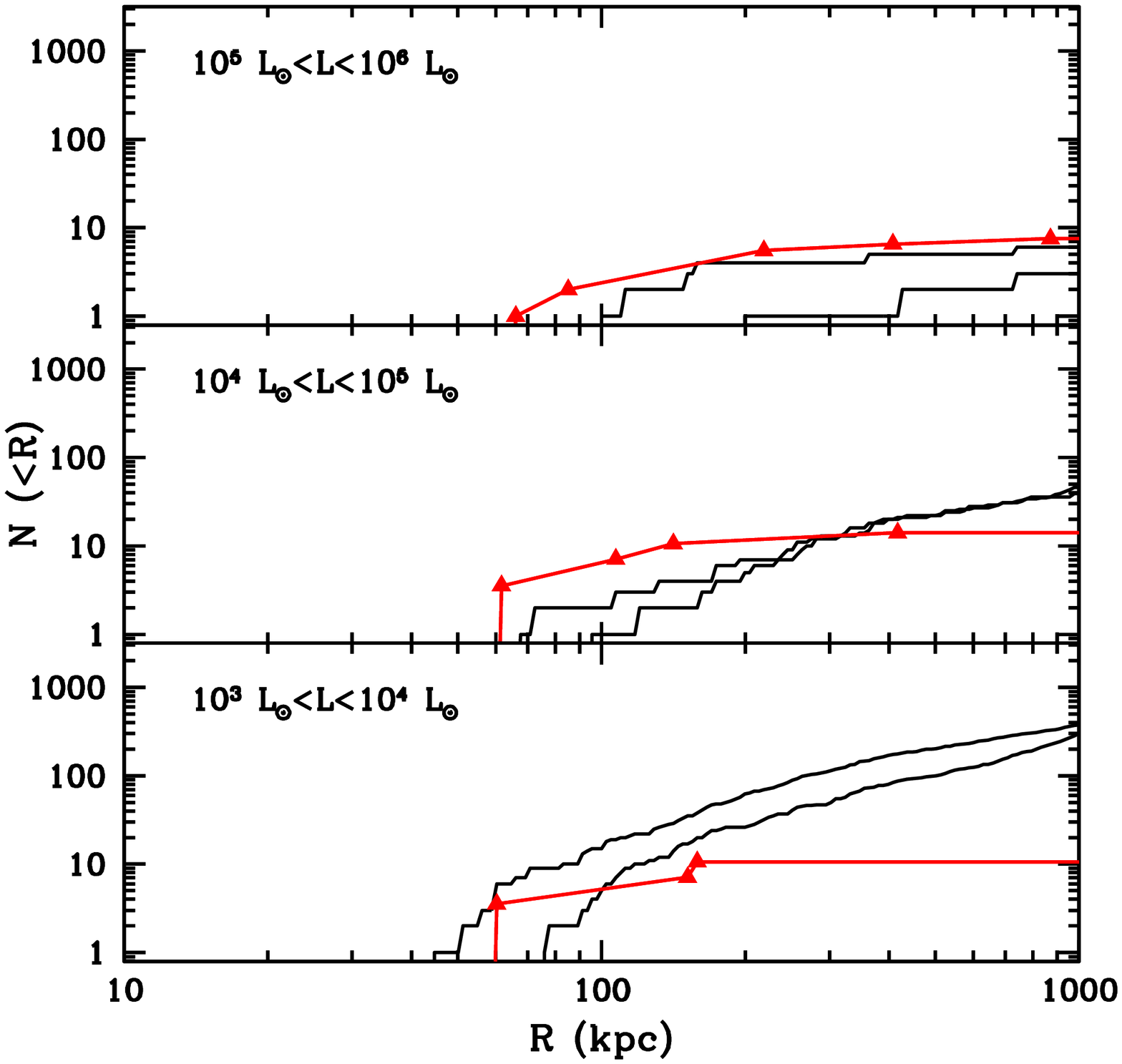}{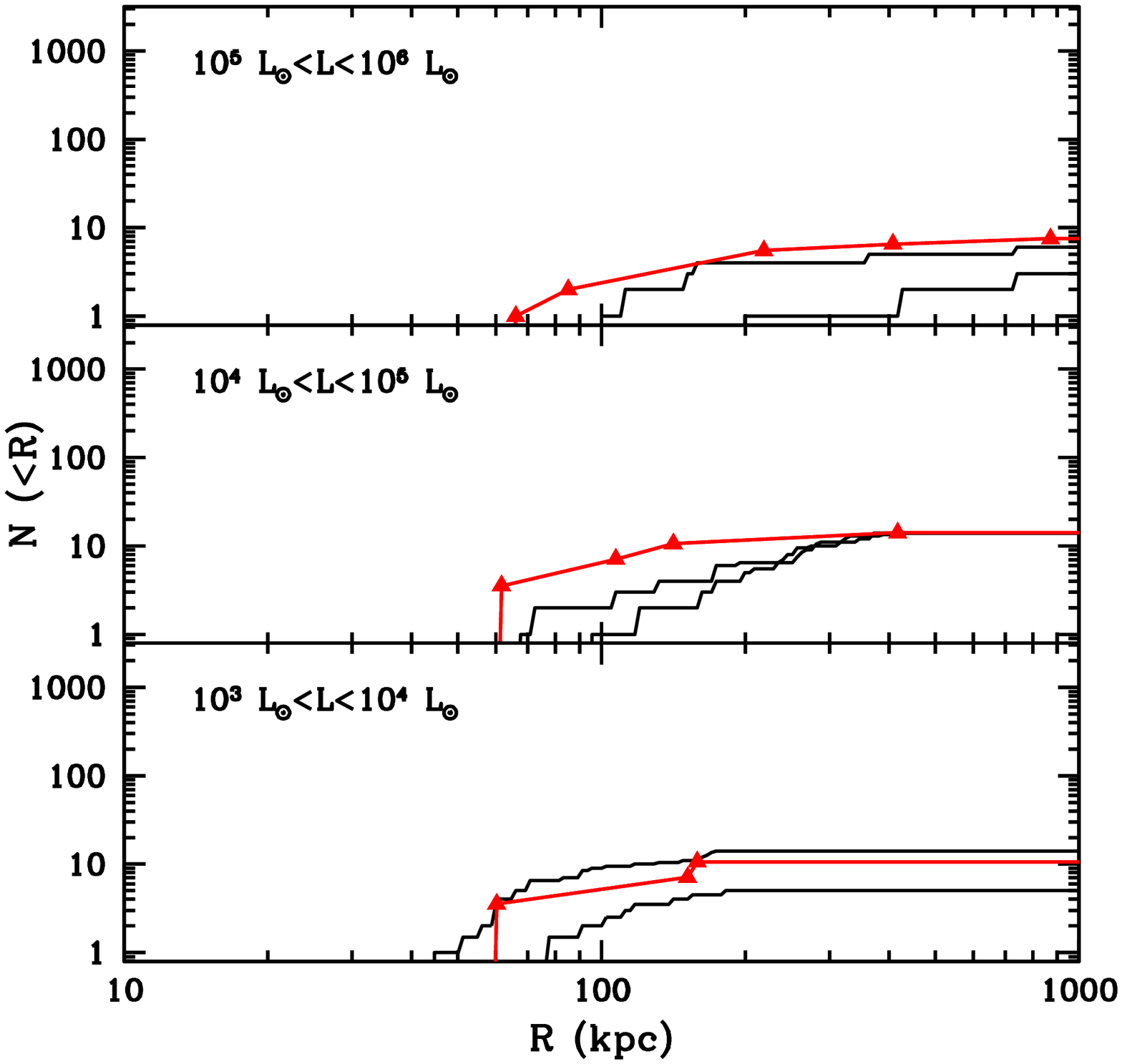}
\caption{{\it{Left}} Same as Fig.~\protect{\ref{RD.all}} but the
  observed satellite distributions only include bona fide fossils: the
  classical dSph which were designated fossils in
  \cite{RicottiGnedin:05} and the ultra-faints whose stellar
  properties match those of the simulated fossil population.  Note
  that this excludes most of the ultra-faints within $50$~kpc. In our
  simulated distributions we use $v_{filter} = 20$~km~s$^{-1}$ to
  define a fossil. We have included all simulated fossils, including
  those which would sit below the SDSS detection limits. {\it{Right}}
  Same as the left panel but simulated radial distributions only
  include the true fossils which would fall within the
  \cite{Walshetal:09} detection limits.}
\label{RD.fos}
\end{figure*}

Figure~\ref{RD.all} shows the galactocentric radial distribution of
all the simulated and observed Milky Way satellites. In the left panel
of Figure~\ref{RD.all}, we compare observations to simulations without
correcting for the sensitivity limits of the SDSS
\citep{Walshetal:09,Koposovetal:07} or whether a satellite is a
fossil. In the right panel, we show all the satellites again, now
applying the \cite{Walshetal:09} limits to the simulated halos around
MW.2 and MW.3. Figure~\ref{RD.fos} shows the galactocentric radial
distribution for only the observed and simulated {\it fossils}. As in
Figure~\ref{RD.all}, the right and left panels show the simulated true
fossils with and without the \cite{Walshetal:09} corrections. The
observational and theoretical fossil definitions are discussed in
Sections~\ref{Obser}~and~\ref{Fossil}, respectively. Our simulations
do not account for tidal stripping of stars, and do not reproduce the
properties of the inner ultra-faint dwarfs, and we do not include them in
Figure~\ref{RD.fos}.

The left panels of Figures~\ref{RD.all} and~\ref{RD.fos} show that at
$L_V \sim 10^5 L_\odot$ the fossils become a significant fraction of
the satellite population, with fossil dominance increasing as
satellite luminosity decreases. This is further illustrated in
Figure~\ref{RD.ratio}, which shows the fraction of subhalos which are
fossils, $N_{fos}/N_{all}$, as a function of distance from the host
for the same luminosity bins as Figure~\ref{RD.fos}, excepting $L_V>10^6 L_\odot$. We find that for
$10^5 L_\odot<L_V<10^6 L_\odot$ bin, the fraction of fossils is
0.05-0.1, with the fraction decreasing as host halo mass
increases. For the lower luminosity bins, $N_{fos}/N_{all}$ converges
to $40-50\%$ and $70-80\%$ for $10^4-10^5 L_\odot$ and $10^3-10^4
L_\odot$ bins, respectively. If we include the inner
ultra-faints in our galactocentric radial distributions, we find a
significant overabundance of observed dwarfs within $50$~kpc of the
Milky Way. The stellar properties of the inner ultra-faint dwarfs do
not agree with the simulated stellar properties of the fossils. As
discussed in Paper~I, and \S~\ref{Obser} of this work, we argue
the majority of these objects may represent a population of tidally
stripped remnants of once more luminous dwarfs. A possible exception,
Segue 1, is discussed in \S~\ref{Obser}. Our simulations are also
unable to reliably resolve $z=0$ halos within $50$~kpc of the Milky
Way. We therefore have excluded anything with $R<50$~kpc from our
comparisons.

\begin{figure*}[t]
\plotone{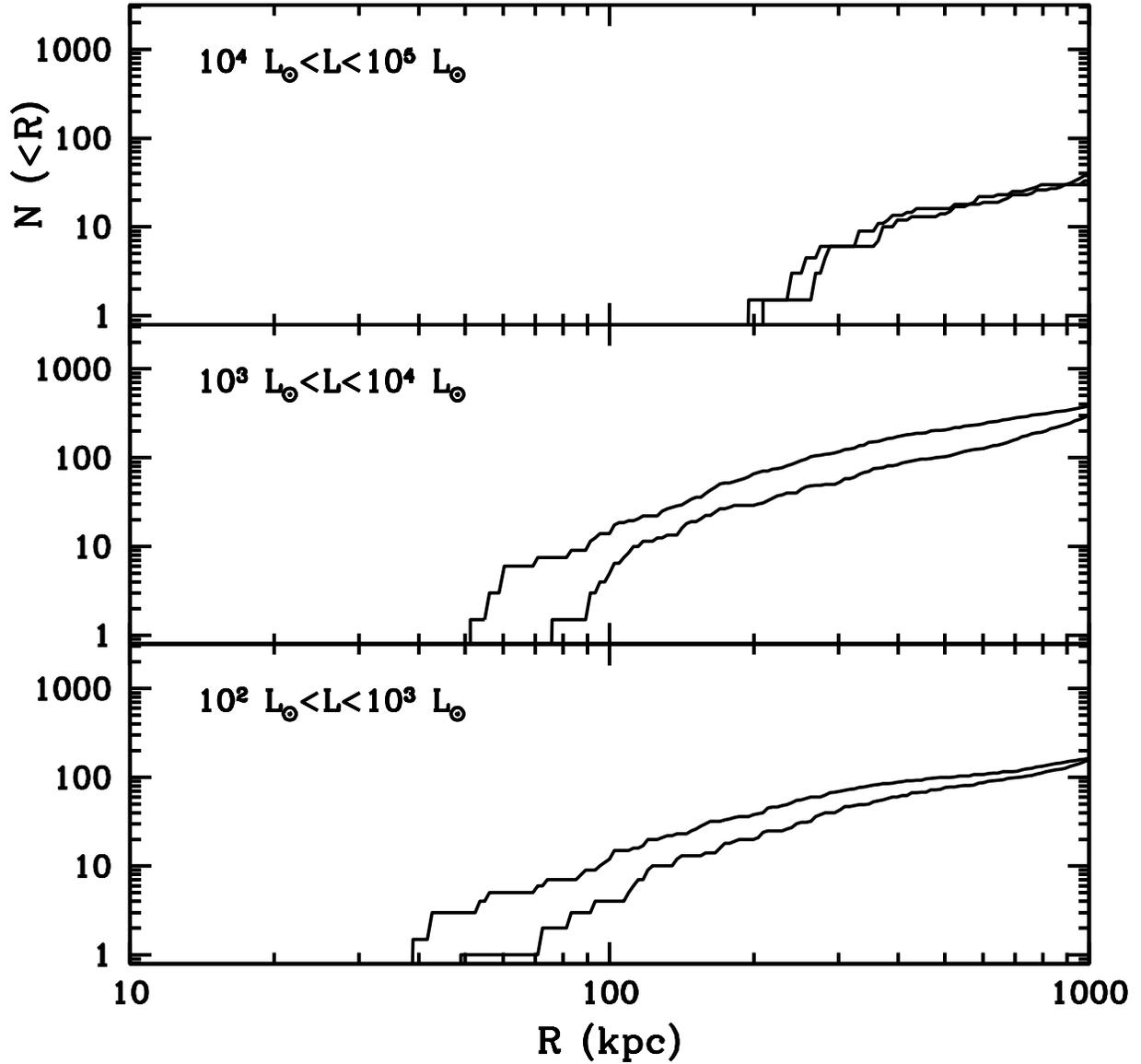}
\caption[Galactocentric radial distribution of the undetected fossils]{The galactocentric radial distribution of the fossils excluding the detectable dwarfs as determined by \cite{Walshetal:09}. Note that the bins have shifted down one order of magnitude in luminosity since there are no undetected fossils with $L_V>10^5 L_\odot$ and we have included the distribution for the lowest luminosity fossils with $L_V < 10^3 L_\odot$.}
\label{RD.undetect}
\end{figure*}

\begin{figure}[t]
\plotone{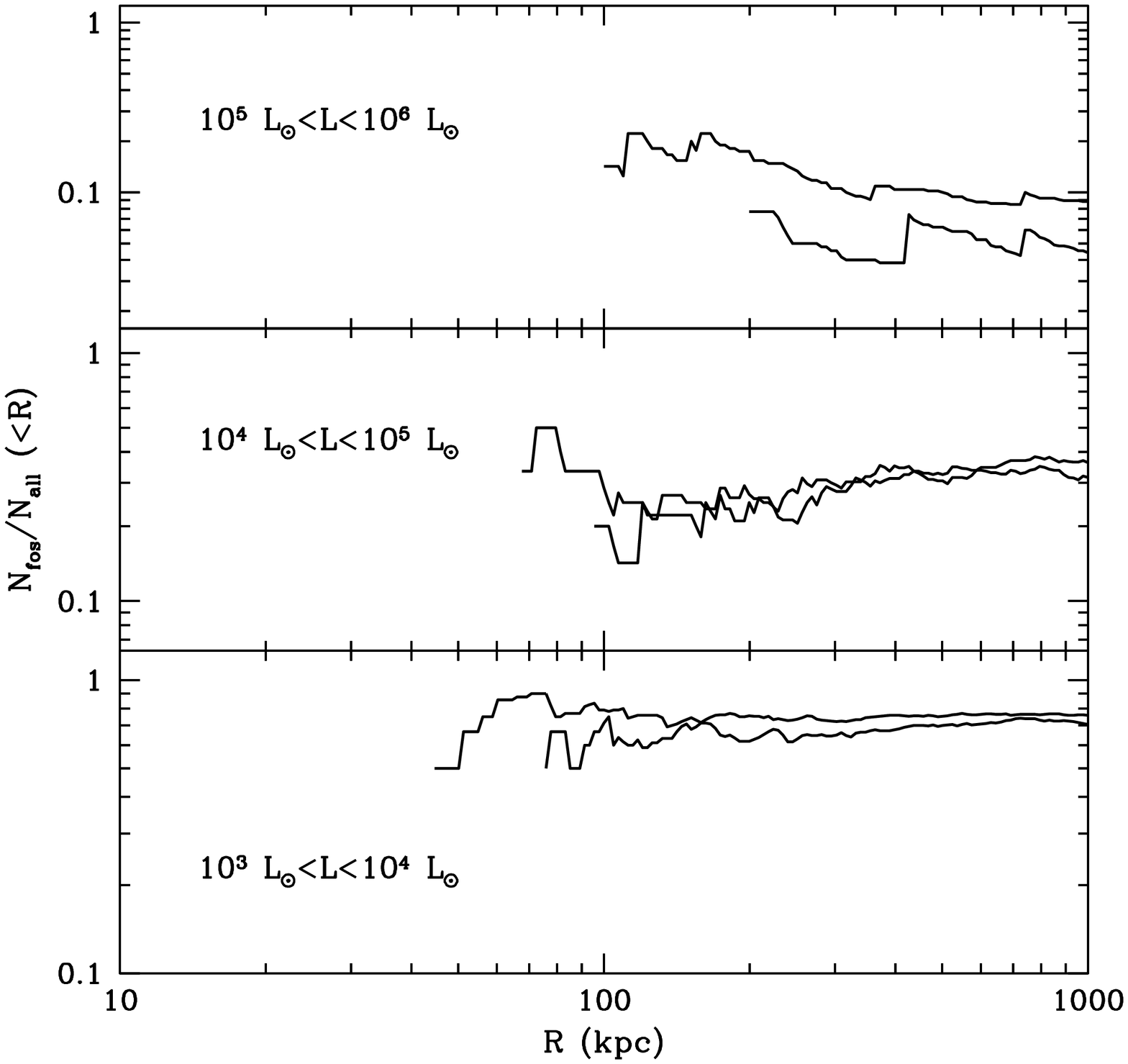}
\caption{The galactocentric radial distribution of the fraction of luminous subhalos which are true fossils around MW.2 and MW.3 from Run D. They are divided in the luminosity bins from Figures~\ref{RD.all}~and~\ref{RD.fos} for which there is a fossil population ($L_V<10^6 L_\odot$).}
\label{RD.ratio}
\end{figure}

Without the ultra-faints with $R<50$~kpc, the right panels of
Figures~\ref{RD.all}~and~\ref{RD.fos} show good agreement between the
simulated satellite distributions of the true fossils around MW.2 and
MW.3 and the observed Milky Way galactocentric radial
distribution. When we convolve our simulated satellite populations
with the limits from \cite{Walshetal:09}, we find that the agreement
between the distribution of dwarfs around MW.2 and MW.3 and that
observed around the Milky Way agree at all radii and luminosity bins
for $L_V < 10^6 L_\odot$ (see right panel of Figure~\ref{RD.fos}). We
thus argue that, in addition to matching the stellar properties of the
ultra-faints, our simulated fossils also agree with their
galactocentric radial distribution.

Figure~\ref{RD.undetect} shows the galactocentric radial distribution of the {\it{undetected}} fossils in our simulations, after excluding detectable fossils according to the detection criterion from \cite{Walshetal:09}. We have not included the bins with $L_V > 10^5 L_\odot$ because there are no undetected fossils in this luminosity range within 1 Mpc of the Milky Way. In addition, all fossils with $L_V > 10^4 L_\odot$ are detected within $200$~kpc. For the lowest luminosity fossils ($L_V < 10^4 L_\odot$) we find $\sim400-500$ undetected dwarfs within 1 Mpc  and $~150$ within $200$~kpc. We have included a panel for very low luminosity bin ($10^2- 10^3 L_\odot$) to look at the distribution of the dimmest fossils which are invisible beyond a few tens of kpc. While the shape of the distribution in the lowest luminosity bins is similar there are approximately two times fewer undetected fossils in the $10^2-10^3 L_\odot$ bin. Given that {\it{fewer}} of the fossils in this bin would be detected compared to its higher luminosity counterpart, we are seeing the decline of star formation in the minihalos with the lowest mass. There are simply fewer $10^2-10^3 L_\odot$ pre-reionization fossils around the Milky Way than their $10^3-10^4 L_\odot$ counterparts.

On the other end of the luminosity spectrum, the right panel of Figure~\ref{RD.all} shows that while our
simulated fossils are able to reproduce the ultra-faint distribution,
we see too many massive, bright ($L_V > 10^4 L_\odot$) satellites at
$R>R_{vir}$, even after the \cite{Walshetal:09} corrections are
applied. We note that this discrepancy does not exist in the lowest, $10^3-10^4 L_\odot$ luminosity bin. This is the first evidence of an apparent discrepancy between
simulations and observations we refer to as the ``bright satellite
problem.'' In the next sections, we will analyze this discrepancy, try
to understand its origin, and whether it can be removed while
maintaining the agreement of the simulations with observations at
smaller radii and lower luminosities.

\subsection{Primordial Cumulative Luminosity Functions}
\label{LFs}

\begin{figure*}
\centering
\plottwo{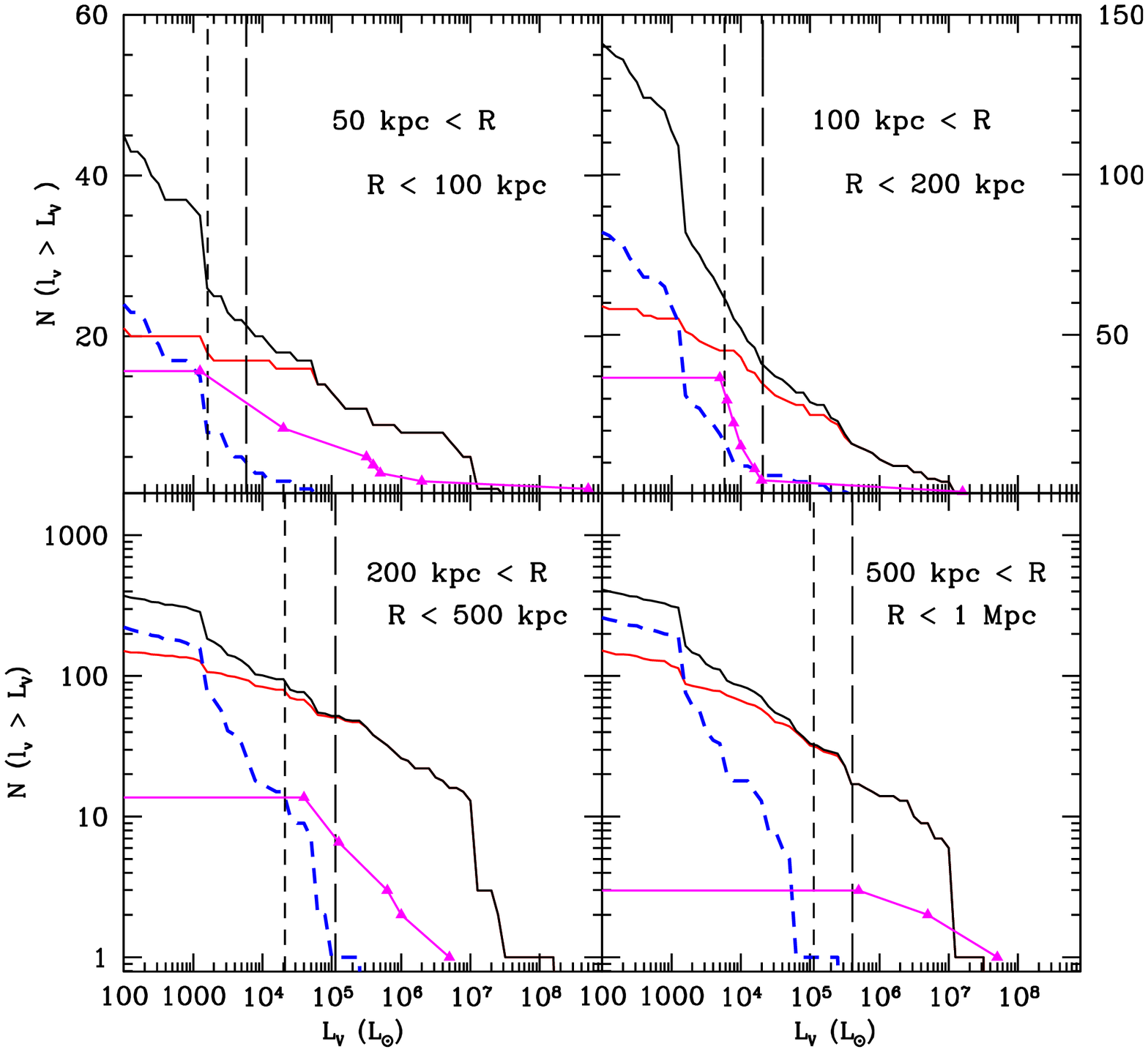}{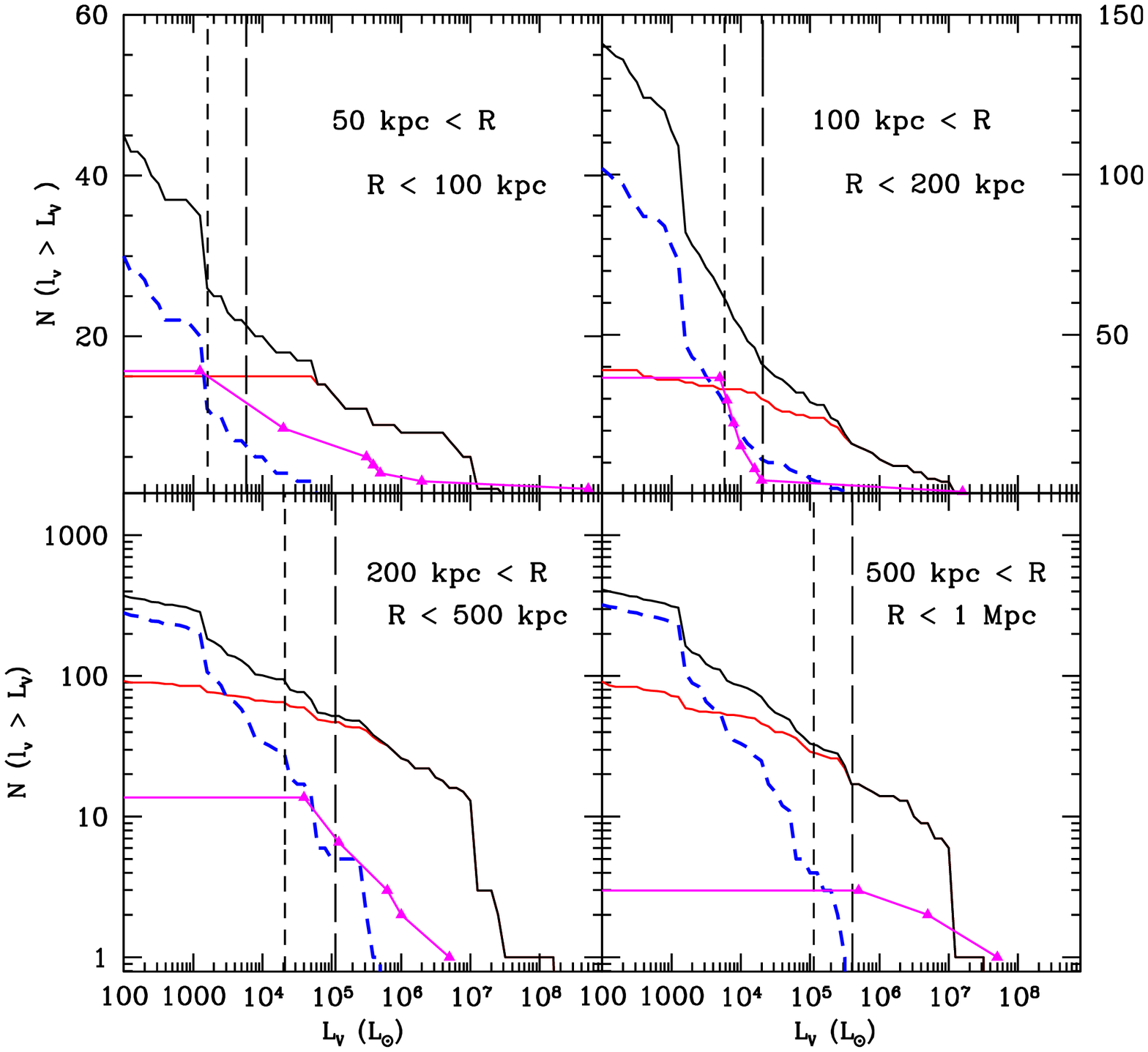}
\caption[width=60m]{({\it Left}). Cumulative primordial luminosity
  function of MW.3 from Run D with the observed luminosity function of
  Milky Way satellites. We have used a $v_{filt}=20$~km~s$^{-1}$ to
  determine whether a simulated halo is a non-fossil or true
  fossil. We show the luminosity functions for four distance bins,
  $50$~kpc$<R<100$~kpc (upper left), $100$~kpc$<R<200$~kpc (upper
  right), $200$~kpc$<R<500$~kpc (lower left), and $500$~kpc$<R<1$~Mpc
  (lower right). In all distance bins the relevant populations are
  noted as follows. The solid black curves show the total cumulative
  luminosity function from our simulations, the red solid lines shows
  the same for only the star forming halos (non-fossils, including polluted
  fossils).  We show the true fossil population with the blue dashed
  curve. The total observed population is shown as magenta triangles with the
  ultra-faint dwarf distribution corrected for sky coverage of the
  SDSS. The detection limits given in \cite{Walshetal:09} are shown as
  vertical black, dashed lines. Long dashed for the luminosity limit
  for the outer radii and shorter dashes for the inner radii in a
  given bin. ({\it Right}). Same a the figure on the left but for
  $v_{filt}=30$~km~s$^{-1}$.}
\label{LF.pristine}
\end{figure*}

\begin{figure}
\plotone{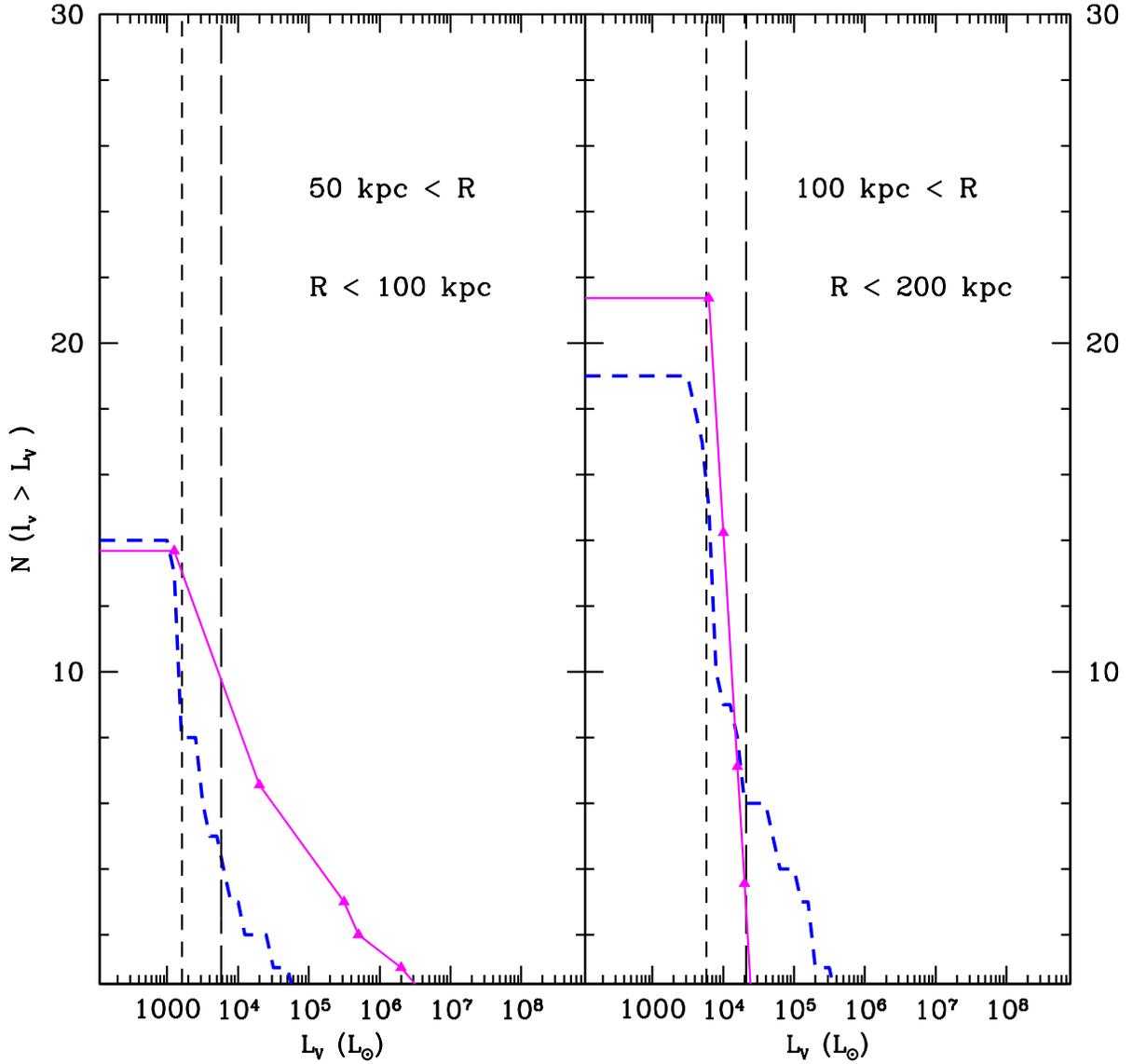}
\caption{Cumulative primordial luminosity function for MW.3 in
  Run D and the total observed population with only the fossils plotted. The simulated
  fossils are shown as the blue dashed line and the observed fossils as
  the magenta triangles. The fossil criteria for the observed
  satellites is the same as in Figure~\ref{RD.fos}. We have only shown
  subhalos around MW.3 which would be detectable by
  SDSS.}
\label{LF.fossil}
\end{figure}

\begin{figure}
\plotone{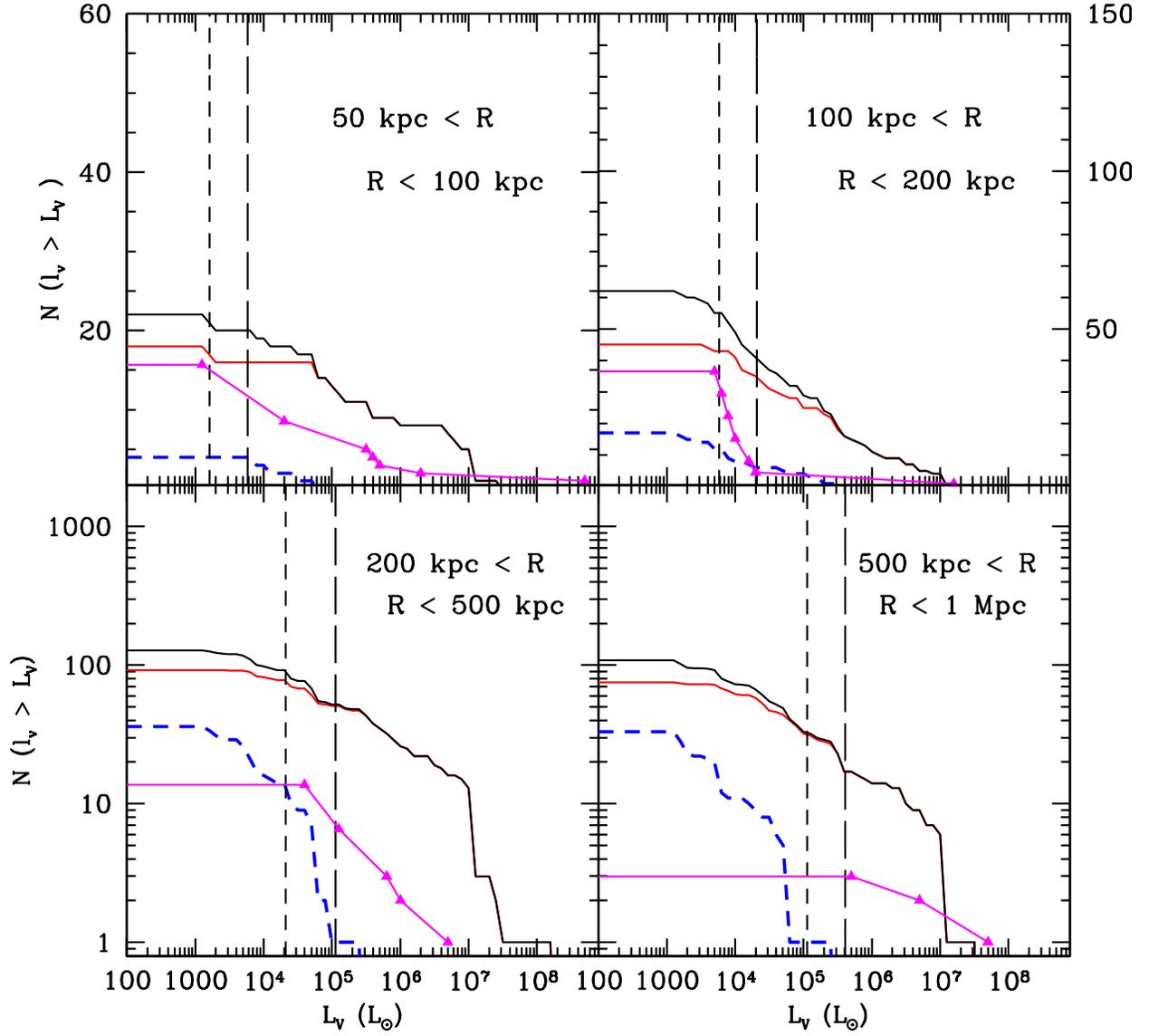}
\caption{Cumulative primordial luminosity function for MW.3 in Run D and
  observations. Only
  simulated dwarfs which have $\Sigma_V$ above the
  \cite{Koposovetal:07} limit are shown.}
\label{LF.Koposov}
\end{figure}

\begin{figure}
\plotone{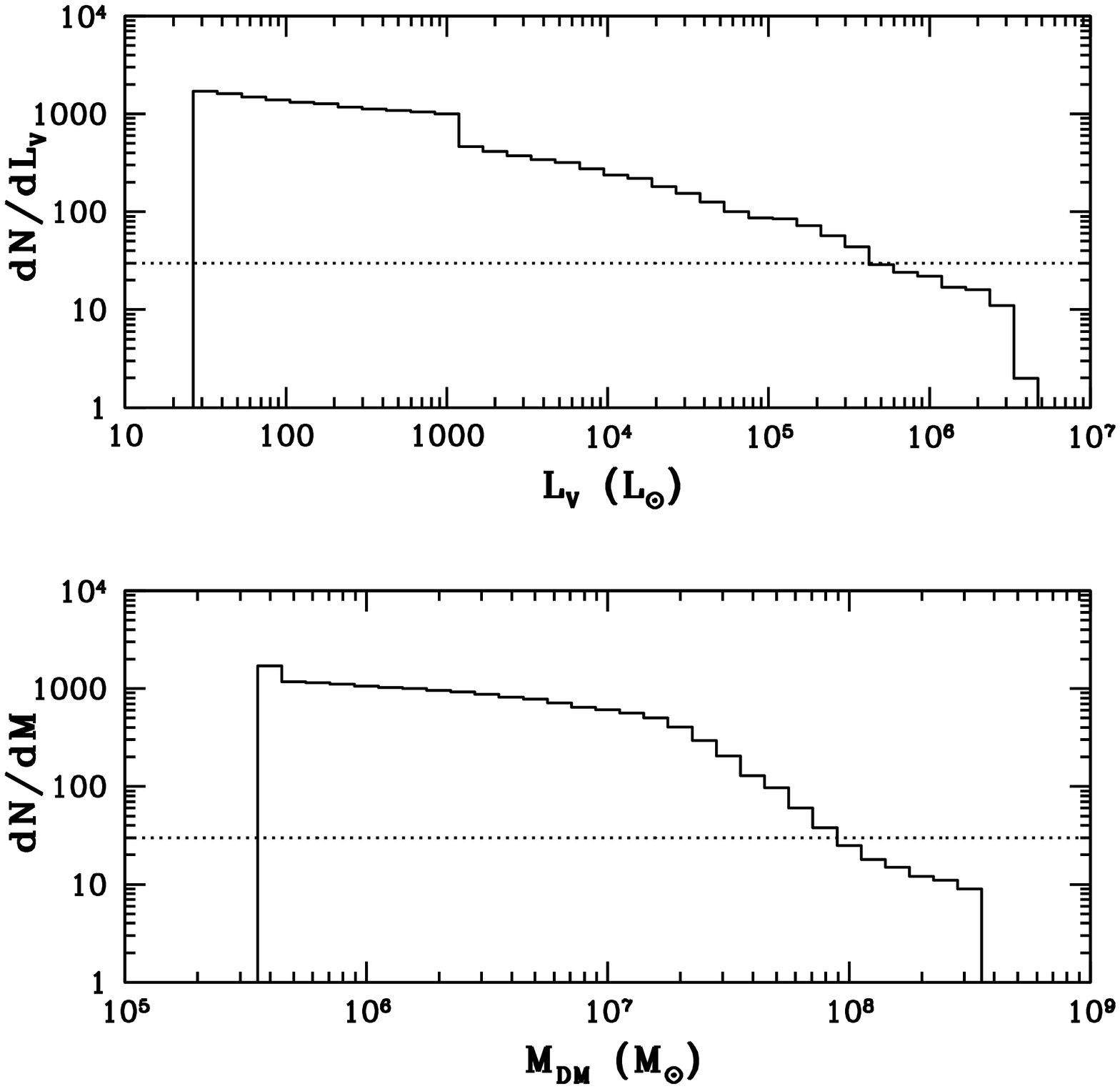}
\caption{The luminosity (top) and mass (bottom) functions for
  pre-reionization halos within $R_{vir}$ of MW.3 which are {\it{not}}
  part of a bound subhalo at $z=0$. We only include those unbound
  luminous pre-reionization halos between $20$~kpc and $50$~kpc, where
  all but two of the known tidal ultra-faints are located. The
  horizontal dotted lines show the approximate number of stripped
  pre-reionization halos required to reproduce the inner ultra-faint population, $\sim 30$.}
\label{LF.MF.part}
\end{figure}

We next explore the fossil distribution and ``bright satellite
problem'' from another angle via comparisons between simulated
cumulative primordial luminosity functions and the observed cumulative
luminosity function at different galactocentric distances from the
Milky Way center. Results are equivalent for all three simulated Milky
Ways and for both versions of our initial conditions. Therefore, for
the remainder of this section and the next we will be discussing the
results for MW.3 in Run D.

Each cumulative luminosity function in this paper is split into four
radial bins to probe different regimes. We choose not to include the
sample at $R<50$~kpc, where the observational sample is the most
complete, because tidal effects are prevalent and our simulations do
not have sufficient resolution to determine whether or not
pre-reionization halos stripped of their enveloping cloud of tracer
particles have been tidally disrupted. The first bin we consider shows
$50$~kpc~$<R<100$~kpc. The next bin out, the outer portion of the Milky
Way halo from $100$~kpc to $200$~kpc, has observations which are
fairly complete for $L_V > 10^4$~L$_\odot$, including the brightest
ultra-faints. From $200$~kpc to $500$~kpc all but two of the
ultra-faints (CVnI and Leo~T) would be below the detection limits of
the surveys and would not be visible.  Roughly, this region
corresponds to the virial radius ($R_{200} \sim 200$~kpc) to $R_{50}$ for a
Milky Way mass halo. Specific to our Local Group, this is the regime
where M31 begins to play a significant role in the satellite counts,
increasing the care required to separate the Milky Way and Local Group
dwarfs from those bound to M31. The final radial bin, from $500$~kpc
to $1$~Mpc, probes the transition region from the edge of the Milky
Way halo to the surrounding filament and void. Subhalos at these radii
are just beginning to fall into the host system, and all the
ultra-faints are below detection limits.

We divide the simulated satellites into fossils and non-fossils:
Figure~\ref{LF.pristine} shows the primordial cumulative luminosity
functions in the four radial bins for fossil thresholds
$v_{filt}=20$~km/s (left panel) and $v_{filt}=30$~km/s (right
panel). The observed cumulative luminosity function is shown as
magenta lines and includes all the classical dwarfs and the
ultra-faints, excepting the population at $R<50$~kpc. The simulated
non-fossils are shown as the red solid curve, and for all bins they
dominate for $L_V > 10^4-10^5$~L$_\odot$. These halos may have been
able to accrete gas and form stars after reionization, and their
primordial cumulative luminosities represent a lower limit for their
present day luminosity. If we were to allow for additional star
formation after reionization, the total number of luminous non-fossils
would remain constant, but the curve would shift to higher
luminosities (to the right). The primordial cumulative luminosity
function of the true fossils (blue dashed curve) has no such
caveat. Their luminosities are known since they have not undergone
post-reionization baryonic evolution aside from the aging of their stellar
populations. The primordial cumulative luminosity function of the
entire simulated population is the solid black curve. Note, that for
$L_V < 10^5 L_\odot$, the total primordial cumulative luminosity
function is increasingly dominated by fossils.

Before looking at the primordial cumulative luminosity functions in
detail, we insure we are comparing equivalent populations. By
definition, all observed Milky Way satellites are above current
detection limits, however, as seen in Paper I, a subset of our
simulated fossils have surface brightnesses below the detection limit
of the SDSS. We use both the \cite{Walshetal:09} and
\cite{Koposovetal:07} limits to test the distribution of detectable
true fossils against observations. Figure~\ref{LF.fossil} shows the
true fossil luminosity function convolved with the \cite{Walshetal:09}
limits (blue dashed line) and the observed fossil sample (magenta
line) used in Figure~\ref{RD.fos}. As in the galactocentric radial
distributions, we find good agreement between the primordial
luminosity function of the true fossils and the observed fossils for
$50-100$~kpc and $100-200$~kpc. We do not make comparisons at
$R>200$~kpc because of the inability of current surveys to detect
fossils at these larger distances. In Figure~\ref{LF.Koposov}, we next
use the surface brightness limits from \cite{Koposovetal:07} to remove
any simulated fossil satellite not detectable by current
surveys. Using the Koposov et al straight surface brightness cut all
but eliminates the fossil population for $v_{filt} =
20$~km~s$^{-1}$. This is a much stronger effect on the detectability
of our true fossils than that seen for the Walsh et al. luminosity and
distance cuts.

In all distance bins, there is an overabundance of the bright
satellites at luminosities typical of the classical dwarfs ($L_V >
10^5 L_\odot$). These should be easily detectable by the SDSS
according to \cite{Walshetal:09} (assuming the undetected dwarfs
have the same distribution of half light radii as the
ultra-faints). In Figure~\ref{LF.pristine}, the detectable dwarfs are
to the right of the dashed line. During our discussion of the missing
bright satellites, we use cumulative luminosity functions which have
not been corrected for the SDSS limits. We now look at each distance
bin individually.

\subsubsection{Inner Ultra-Faint Dwarfs at $R<50$~kpc}

In Paper I, we argue that, while the inner ultra-faints have likely
lost significant fractions of their stellar populations to tidal
striping, they were not necessarily dIrr at the start of their
encounters with the Milky Way. Instead, they may have been more
massive primordial fossils. We base this conjecture on Figure~12 in
Paper I which shows that the inner ultra-faints have metallicities,
[Fe/H], that are similar to fossils that are slightly more
luminous. But are there enough massive fossils to account for the
inner ultra-faints? Figure~\ref{LF.MF.part} shows the mass function
(bottom) and luminosity function (top) of the pre-reionization halos
which are not part of a bound halo at $z=0$ and are between $20$~kpc
and $50$~kpc from MW.3. The dotted horizontal lines show the
approximate number of stripped fossils required to reproduce the inner
ultra-faints. We see that to produce the $\sim 30$ inner ultra-faints
around the Milky Way, we would only need to consider the largest
primordial fossils with masses at reionization $M>10^8$ ~M$_\odot$ and initial
luminosities $L_V>10^6$~L$_\odot$.

\subsubsection{$50~kpc<R<100$~kpc}

A strong piece of evidence for the primordial model would be the total
number of observed satellites in one or more radial bins being greater
than the number of non-fossils. When we look at $R<100$~kpc without
the $R<50$~kpc cut, we see such an overabundance of observed
dwarfs. However, when we do not include the dwarfs within $50$~kpc of
the galactic center the case is no longer clear cut. If the satellite
count from $50-100$~kpc increases to greater than 25, there is a case
for fossils even using the most conservative
$v_{filt}=20$~km~s$^{-1}$. For $v_{filt}=30$~km~s$^{-1}$ the number of
non-fossils available from $50-100$~kpc drops to $~\sim18$.

For luminosities at which the observational sample is complete, to the
right of the dashed lines, we see too many bright ($L_V > 10^4
L_\odot$) objects, even in this inner most radial bin. In addition, as
our simulations do not account for post-reionization star formation,
it is likely that the overabundance of bright objects is worse than
shown in Figure~\ref{LF.pristine}. Unless all of the non-fossils have
accreted no gas and formed no additional stars after reionization, the
simulated curve must lie {\it{below}} the observations. This allows
these star forming halos to increase in luminosity, shifting the
luminosity function of the non-fossils to the right.

\subsubsection{$100~kpc~<R<200$~kpc}

In this bin, we probe the outer reaches of the Milky Way's virial
halo and there are a few notable characteristics of the luminosity
functions. First, with the addition of the observed dwarfs in this
bin, the total number of known satellites around the Milky Way
increases to $\sim 65$, $\sim 45$ not including the inner
ultra-faints. The sample at these larger radii is only complete for
$L_V > 10^4$~L$_\odot$. For $100-200$~kpc, with a
$v_{filt}=20$~km~s$^{-1}$ and the less conservative
$v_{filt}=30$~km~s$^{-1}$ there are $\sim 60$ and $\sim 40$ simulated
non-fossils respectively. Second, the presence of undetected dwarfs is
corroborated by the shape of the observed luminosity function around
$10^4$~L$_\odot$. Not only is it rising steeply to the detection limit
at $R=100$~kpc, but its shape is similar to the simulated primordial
luminosity function for the true fossils.

In the outer virial halo, we once again overproduce the number of
bright satellites. At these radii the discrepancy between theory and
observation is more severe than for $50$~kpc$<R<100$~kpc since for
these radii there is only one observed satellite with $L_V > 3
\times10^4$~L$_\odot$.

\subsubsection{$R>200$~kpc}

For $R>200$~kpc, we can only make observational comparisons for $L_V > 10^5$~L$_\odot$. Beyond the virial radius the discrepancy between the observed number of satellites and our simulations is up to $\sim1.5$ orders of magnitude, compared with factors of $\sim 2$ and $\sim 10$ for the $50~{\rm  kpc}<R<100$~kpc and $100$~kpc$<R<200$~kpc bins, respectively.

\section{Where are the Bright Satellites?}
\label{LuSP}

Our simulations of the fossils of the first galaxies are consistent
with the observed Milky Way satellite galactocentric radial
distributions (Section~\ref{RDs}), primordial cumulative luminosity
functions (Figure~\ref{LF.fossil}), and internal stellar properties (see Paper I). More intriguing, is that at first glance the
model appears to fail in the outer parts of the Milky Way by
over-producing the number of bright non-fossil satellites. In this
section, we explore possible solutions to the bright satellite problem
and whether the proposed solutions maintain the agreement between the
observed ultra-faint dwarf population and our simulated true
fossils. First, we will explore whether the model overrestimates the
star formation efficiency in pre-reionization dwarfs, and, then, whether
we have too many luminous galaxies forming before reionization. It
would be of great interest if we could use current observations to
constrain galaxy formation models before reionization. We conclude the
analysis with a proposal in which the bright satellites do exist in
the outer parts of the Milky Way halo but may still be elusive to
detection due to their extremely low surface brightnesses. The discovery
of these ``ghost halos'' around known isolated dwarfs is a test of the
primordial model and would allow us to determine the efficiency of star formation in
the first galaxies.

\subsection{Increasing Mass to Light Ratios}

\begin{figure}
\centering
\plotone{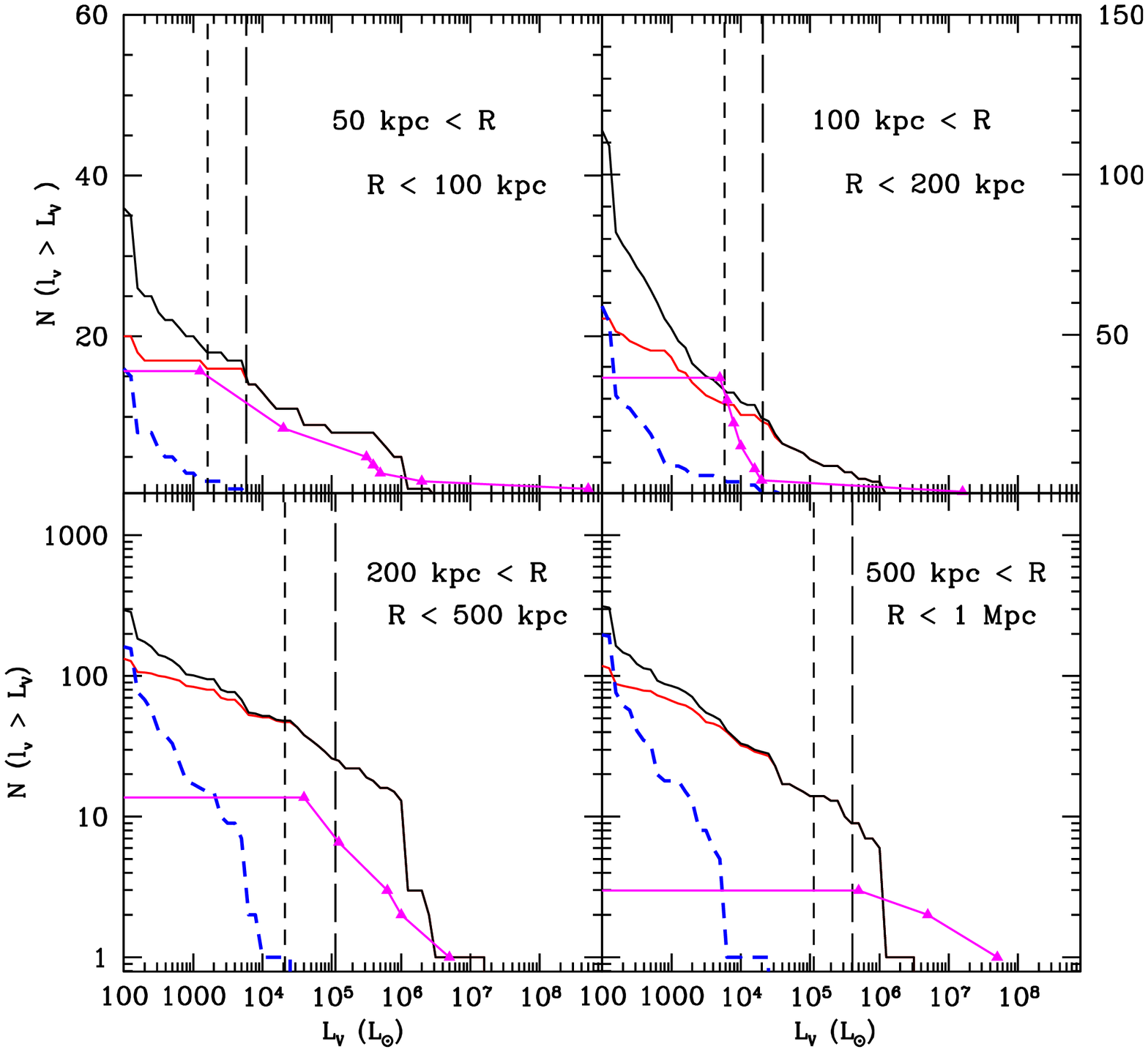}
\caption{Cumulative luminosity functions of MW.3 from Run D (colored
  curves) with the total observed population (magenta triangles). All the symbols and
  lines mean the same as in Figure~\ref{LF.pristine}. Here we increase
  the stellar mass-to-light ratio by a factor of 10 to 50
  $M_\odot/L_\odot$.}
\label{LF.ml50}
\end{figure}

\begin{figure}
\centering
\plotone{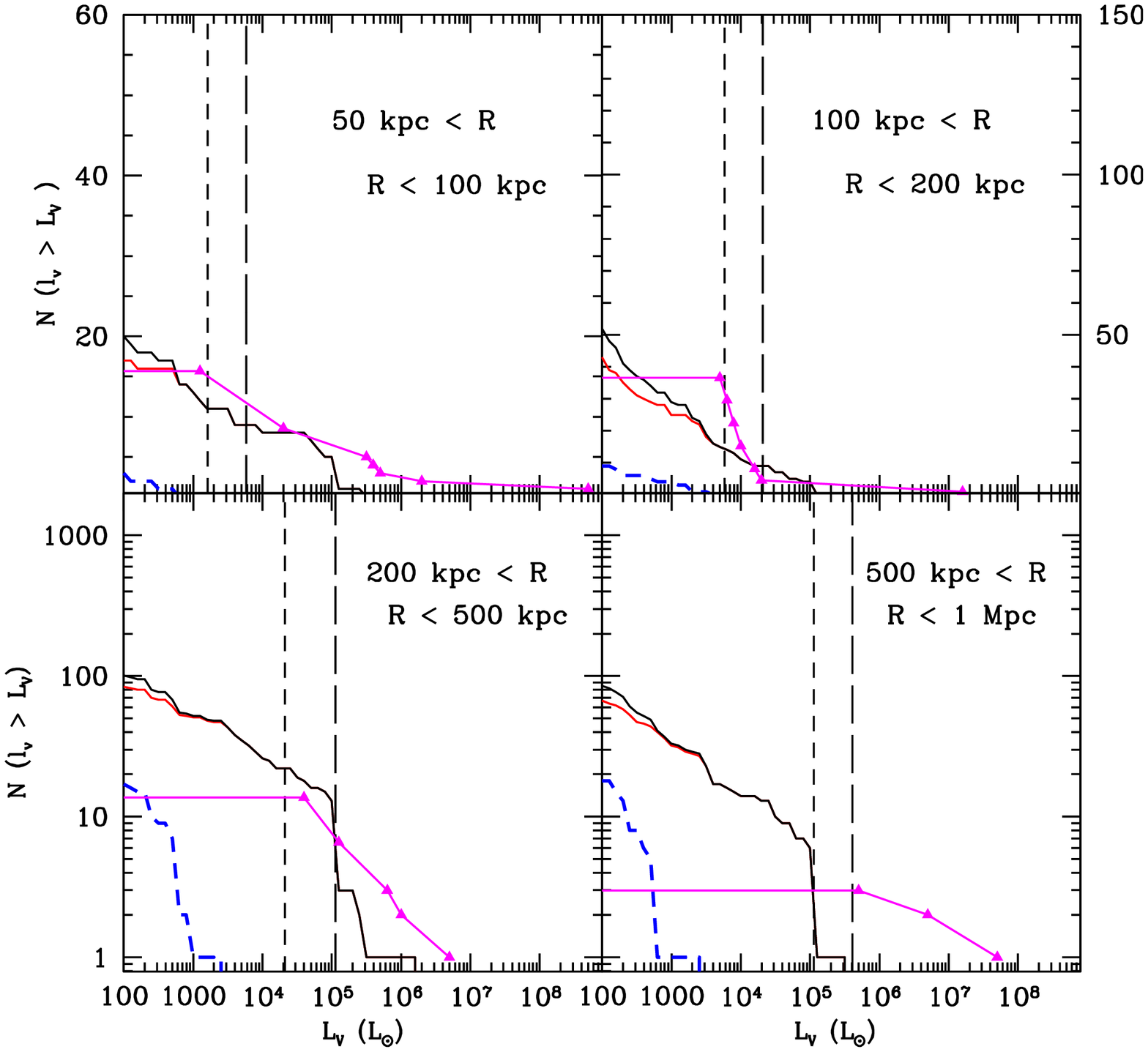}
\caption{Cumulative luminosity functions of MW.3 from Run D (colored
  curves) with the total observed population (magenta triangles). All the symbols and
  lines mean the same as in Figure~\ref{LF.pristine}. Here we increase
  the stellar mass-to-light ratio by a factor of 100 to 500
  $M_\odot/L_\odot$. }
\label{LF.ml500}
\end{figure}

First, we explore whether our pre-reionization simulations overestimate the luminosity of galaxies independently of dark matter halo masses. This could be due to using an incorrect IMF in the pre-reionization simulations. We use a mass-to-light ratio for the aged stellar population,
\begin{equation}
\frac{M_\ast^{rei}}{L} = \left({M_\ast^{today} \over L}\right)\left({M_\ast^{rei} \over M_\ast^{today}}\right)
\end{equation}
where $M_\ast^{rei}$ and is the mass of the stellar population at reionization, and $M_\ast^{today}$ is the mass of the stellar population at $z=0$. The ratio of $M_\ast^{rei}/M_\ast^{today}$ depends on the primordial IMF. The ratio of $M_\ast^{rei}/M_\ast^{today}$ depends on the primordial IMF, and $M_\ast^{today}/L ~1-2$ as for the oldest globular clusters in the Milky Way. A more top heavy IMF for the primordial stars will result in greater stellar mass loss after reionization and fewer low mass stars which can survive to the modern epoch. Conversely, the ratio drops as the primordial IMF produces fewer high mass stars. 

To approximate this effect, in Figure~\ref{LF.ml50}, we plot the cumulative luminosity functions as in
Figure~\ref{LF.pristine}, but increase the stellar mass to light ratio
by a factor of 10 in our pre-reionization dwarfs, to $50
M_\odot/L_\odot$.  The figure shows that increasing the mass to light
ratio to $50 M_\odot/L_\odot$ does not decrease the number of luminous
satellites enough to match observations. In the $50$~kpc~$<R<100$~kpc
bin, we can match observations. However, since the primordial
luminosities of the non-fossils are only lower limits, the agreement
disappears if the population formed {\it{any}} stars after
reionization. If they did, we are still over-producing luminous
satellites.  We need to use a mass to light of $500 M_\odot/L_\odot$
to not over-produce the number of non-fossil satellites in any radial
bin. However, this high mass to light ratio makes the fossils
virtually dark, with $L_V < 10^{2}$~L$_\odot$.

A blanket suppression of star formation in all halos does not solve
the bright satellite problem unless we suppress all star formation in
most halos before reionizatioin.  We next explore suppression
mechanisms which are dependent on the environment or properties of the
halos.

\subsection{Suppression of Pre-Reionization Dwarf Formation in Voids}

\begin{figure}
\centering
\plotone{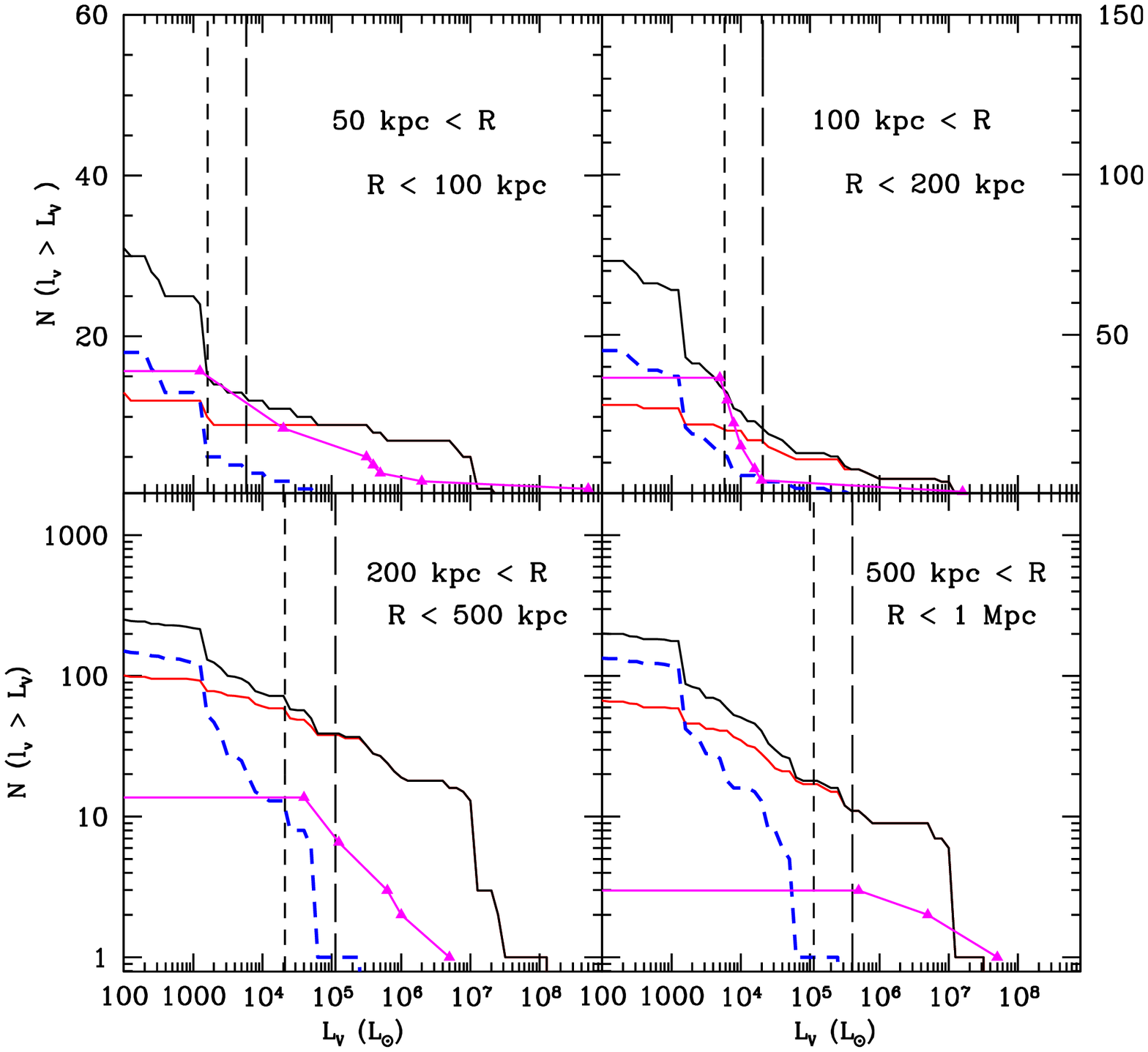}
\caption{Cumulative luminosity functions of MW.3 from Run D (colored
  curves) with the total observed population (magenta triangles). All the symbols and
  lines mean the same as in Figure~\ref{LF.pristine}. Here we have
  completely suppressed star formation in any halo outside the highest
  density regions. This is the most extreme case of lower $H_2$
  formation, and therefore lower star formation in the voids.}
\label{LF.high10}
\end{figure}

The formation of $H_2$ in the early universe is catalyzed by ionizing
UV radiation emitted by nearby star forming regions
\citep{RicottiGnedinShull:02b}. In the voids, two factors work against
$H_2$ formation. The delay of structure formation in the voids
relative to higher density regions will prevent the minihalos from
collapsing until lower redshifts when the $H_2$ dissociating
background is stronger. In addition, the importance of positive
feedback is reduced due to the larger mean distances between minihalos
in the voids and sources of ionizing radiation
\citep{RicottiGnedinShull:02a,RicottiGnedinShull:02b,RicottiGnedinShull:08}. The
combination of these factors may result in a reduced abundance of
$H_2$ relative to the regions around a Milky Way. This may produce a
star formation efficiency before reionization that depends on the
environment.  We approximate the most extreme case of $H_2$
suppression in the voids by suppressing all star formation in halos in
regions with $\delta \simlt 0.4$. The extreme suppression of the star
formation in the voids we use treats all halos but those in the
overdense regions ($z_{eff}=8.3$) as dark.

Since the bright satellite problem is most prominent in the outer
regions of the Milky Way halo, the lack of star formation in low
density regions may decrease the number of bright halos beyond the
virial radius while leaving the satellite luminosity functions
unchanged at smaller radii and lower luminosities.  However,
Figure~\ref{LF.high10} shows that, even in the most extreme case,
suppressing $H_2$ formation in the voids does not decrease the number
of bright satellites enough in any radial bin to bring our simulations
into agreement with observations. In this scenario, there is no
appreciable reduction of the number of bright satellites for
$R<200$~kpc. There is a decrease in the luminosity function at larger
radii, but it is neither strong or focused enough on the high
luminosity subhalos to solve the overabundance of bright satellites in
the outer parts of the Milky Way. Enough of the region within $1$~Mpc
of our Milky Way is at the highest density in our simulation ($z_{eff}
= 8.3$) that the complete suppression of star formation in even
moderately less dense regions does not sufficiently change the
luminosity functions or radial distributions.

With the inability of $H_2$ suppression in the voids and overall
suppression of star formation to account for the missing bright
satellites, we shift our focus to properties of the halos which vary
with halo mass.

\subsection{Lowering the Star Forming Efficiency}

\begin{figure*}
\centering
\plottwo{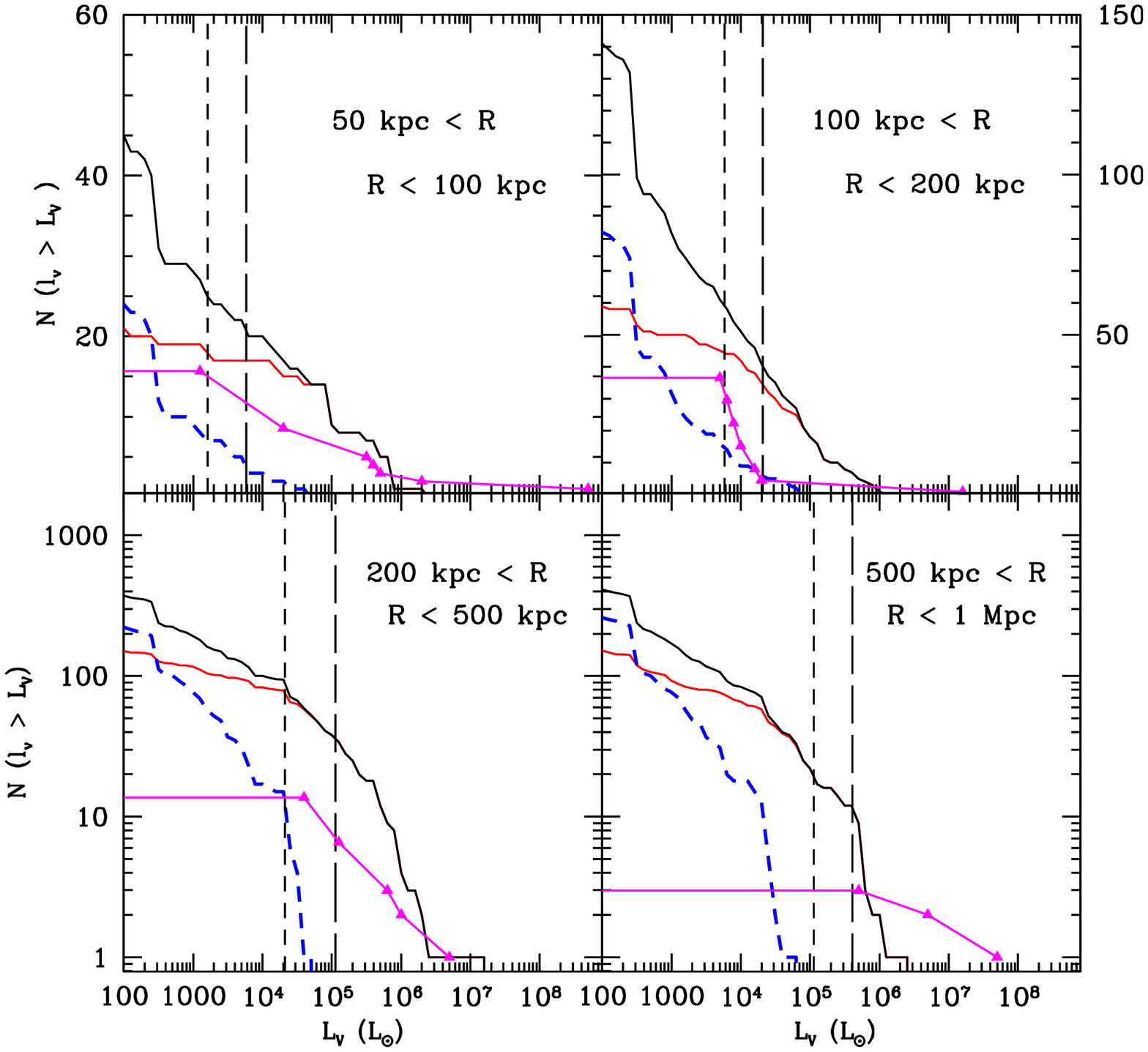}{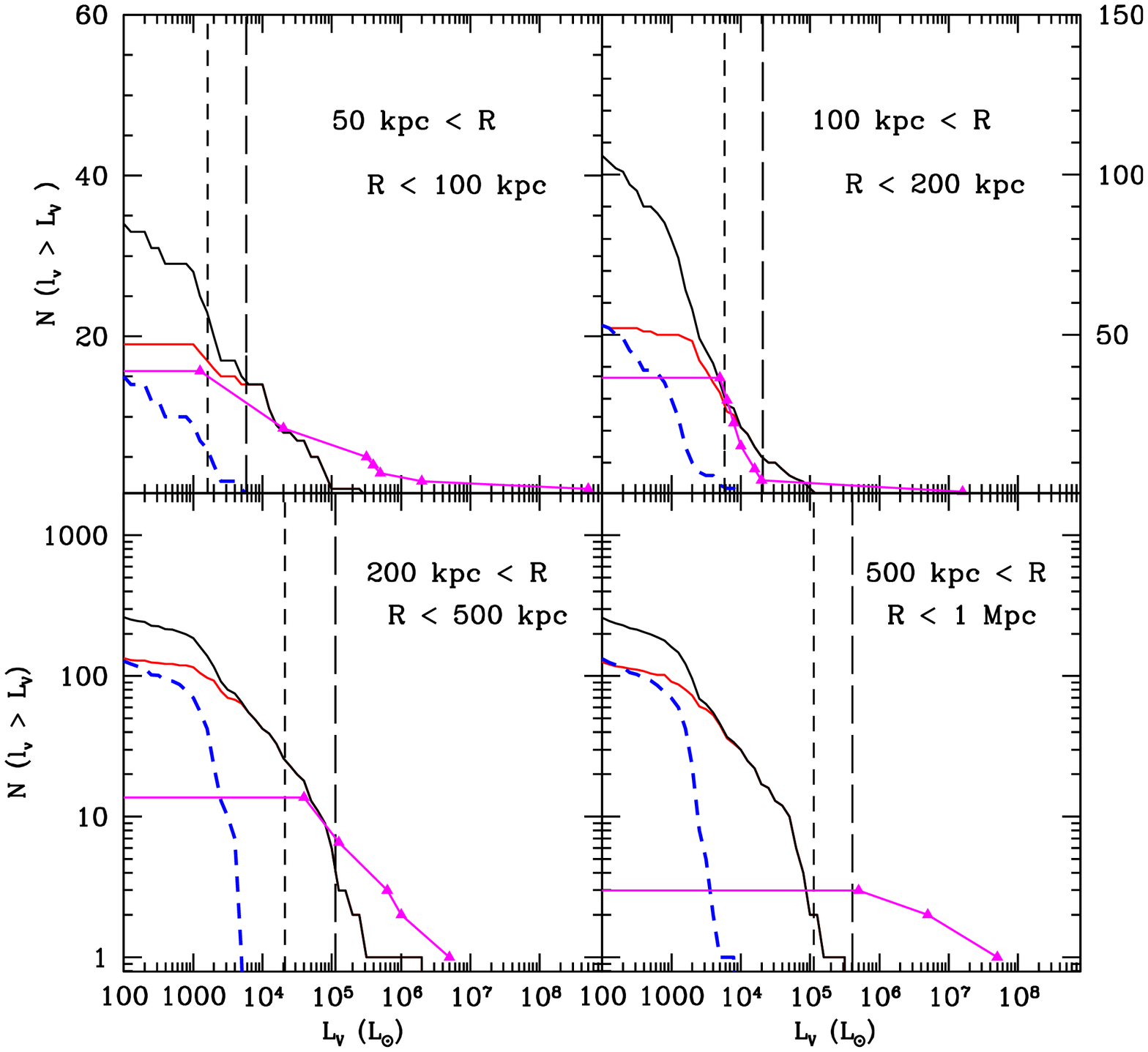}
\caption{({\it Left}). Cumulative luminosity function of MW.3 from Run D (colored curves) and the total observed population (magenta triangles).  All symbols and lines mean the same as in Figure~\ref{LF.pristine}. We have applied an $f_{crit}=1\%$. ({\it Right}). Same as left panel but for $f_{crit}=0.1\%$.}
\label{LF.sup}
\end{figure*}

\begin{figure*}
\plottwo{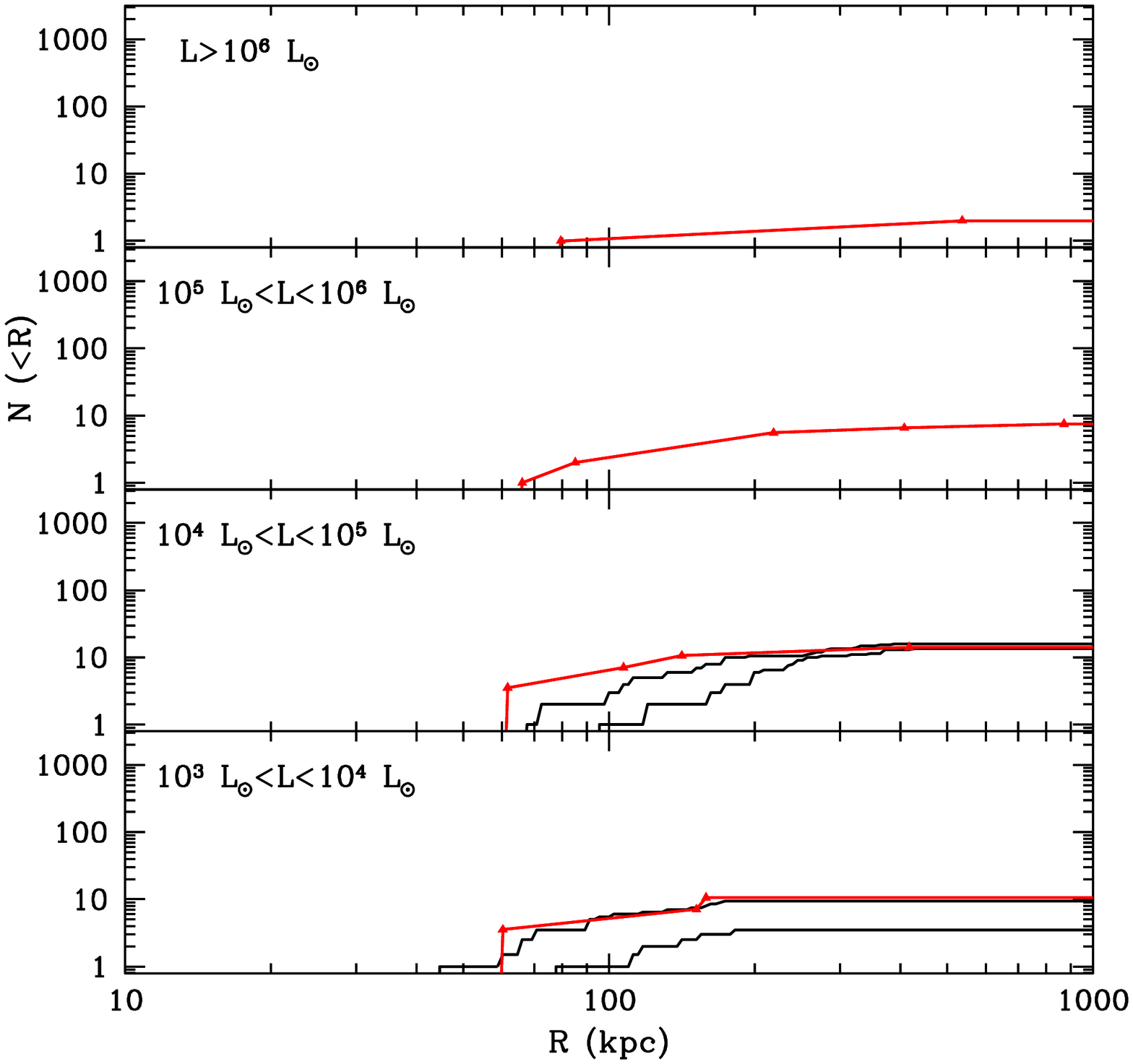}{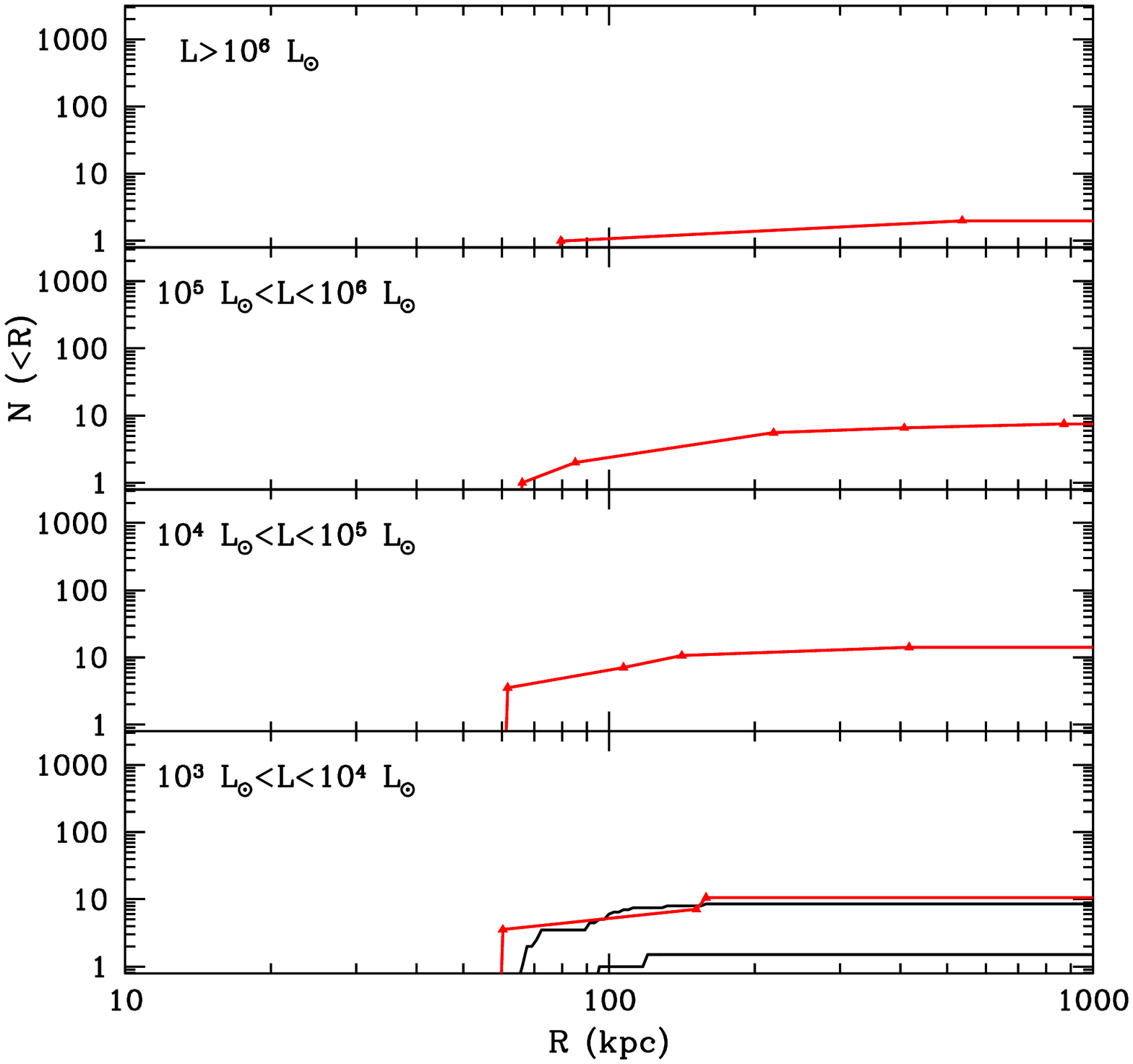}
\caption{({\it{Left}}) : Radial distribution for our simulated MW.2 and MW.3 with a $f_{crit}=1\%$ (black solid curves) and for the fossil Milky Way satellites (red triangles).  Note, that while we are still able to reproduce the radial distribution of the lowest luminosity bin, we are no longer able to match the fossil population for $L_V > 10^5$~ L$_\odot$. ({\it{Right}}) : Radial distribution for $f_{crit}=0.1\%$. Note that for the more extreme suppression, we have lost the fossils population with  $L_V > 10^4$~ L$_\odot$}
\label{RD.Xsup.fos}
\end{figure*}

In the pre-reionization simulations, the sub-grid recipe for star
formation depends on a free parameter, $\epsilon_\ast$, controlling
the efficiency of conversion of gas into stars per unit dynamical
time. One of the main results of the pre-reionization simulations is
that the global star formation rate and $f_*(M)=M_*/M_{bar}$ (the
fraction of total mass converted into stars, where $M_{bar}=M/7.5$) is
nearly independent of $\epsilon_\ast$ in small mass dwarfs due to the
self-regulation mechanisms of star formation.  However, in halos with
masses $M \simgt 5-10 \times 10^7$~M$_\odot$, $f_*(M)$ is typically
proportional to $\epsilon_\ast$ since the higher mass minihalos are
less sensitive to self-regulation feedback. We used $\epsilon_\ast
=5\%$ in our fiducial runs, that may be too large
\citep[\eg,][]{Trentietal:10}. The pre-reionization simulations may have
overestimated the luminosity of primordial dwarfs with $M \simgt 5
\times 10^7$~M$_\odot$, for which the $f_\ast$ vs $M$ relationship is
tighter. We explore the effect of reducing $\epsilon_\ast$ by
introducing a maximum stellar fraction, $f_{*,crit}$. Roughly,
$f_{\ast,crit}$ corresponds to the mass threshold where feedback
effects no longer dominate and where the value of $\epsilon_\ast$
becomes important. If we reduce $f_{*, crit}$, we will decrease the
luminosities of our most luminous halos.  Roughly, for halo masses $M
\sim 3 \times 10^7$~ M$_\odot$ (virial mass at formation) our
simulations have $f_{*}(M) \sim 1\%$. This is in agreement with
observed values for dwarfs with $v_c \sim 10$~km~s$^{-1}$
\citep{McGaughetal:10}.

Figure~\ref{LF.sup} shows the luminosity functions of our simulations
with all halos with $f_{\ast, crit}=1\%$ (left panel) and $f_{\ast,
  crit}=0.1\%$ (right panel). The figure shows that lowering the star
formation efficiency preferentially for the higher mass halos is
effective in decreasing the number of non-fossil subhalos with $L_V >
10^5$~L$_\odot$.  Adopting $f_{*,crit}=1\%$ decreases the number of
luminous halos enough to bring the luminosity functions in agreement
with observations, while preserving the agreement for the fossil
population. However, in the radial bins $50$~kpc~$<R<500$~kpc,
there still are too many subhalos with $L_V > 10^5$~L$_\odot$, though the discrepancy has dropped
significantly.  Coupling a lower $\epsilon_\ast$ with a higher mass to
light ratio or $H_2$ suppression in the voids does not correct the
remaining bright satellite overabundance.  However, when we set $f_{*,crit}=0.1\%$, the cumulative primordial luminosity function of
our simulated dwarfs becomes consistent with observations at all
radii, but requires a deduction of $f_\ast$ in halos with mass $M \sim
7 \times 10^6$~M$_\odot$ whose $f_\ast$ is self-regulated and thus
independent of $\epsilon_\ast$ \citep{RicottiGnedinShull:02b}.

Any solution for the overabundance of bright satellites must preserve
not only the existence of the true fossil population, but also its
distribution and properties. We next look at the other dimensions of
the agreement between the true fossil populations with an
$f_{crit}=1\%$~and~$0.1\%$ and the ultra-faints and classical
dSph. Figure~\ref{RD.Xsup.fos} shows the radial distribution of the
true fossils around MW.2 and MW.3 from Run D the observed Milky Way
population for an $f_{crit}=1\%$. While for $10^3 L_\odot < L_V <
10^4$~L$_\odot$ and $10^4 L_\odot < L_V < 10^5$~L$_\odot$ the fossil
population with $f_{crit}=1\%$ reproduces the observed radial
distribution, we no longer have any true fossils with $L_V>10^5
L_\odot$.  If the star formation efficiency of our pre-reionization
halos is lowered enough to bring the number of luminous satellites in
line with observations, the fossil luminosity threshold discussed in
Paper I is dropped to $L_V < 10^5 L_\odot$. For $f_{crit}=0.1\%$ the threshold drops further to $10^4 L_\odot$.

The loss of the multi-dimensional agreement between our true fossils
and the ultra-faints shows that lowering $\epsilon_\ast$ enough to
account for the missing luminous satellites is not a viable solution
for the bright satellite problem if the ultra-faint dwarfs are fossils
of the first galaxies. In this interpretation,
we need a different mechanism which will either preferentially
suppress star formation in the most luminous pre-reionization halos to
a greater degree, or cause their lower redshift counterparts to lose
the majority of their primordial stellar population after
reionization. If neither of these solutions work, we must ask ourselves if the halos CDM predicts are there at all.

\subsection{The Ghost Halos}
\label{SEC.gh}

\begin{figure*}
\plottwo{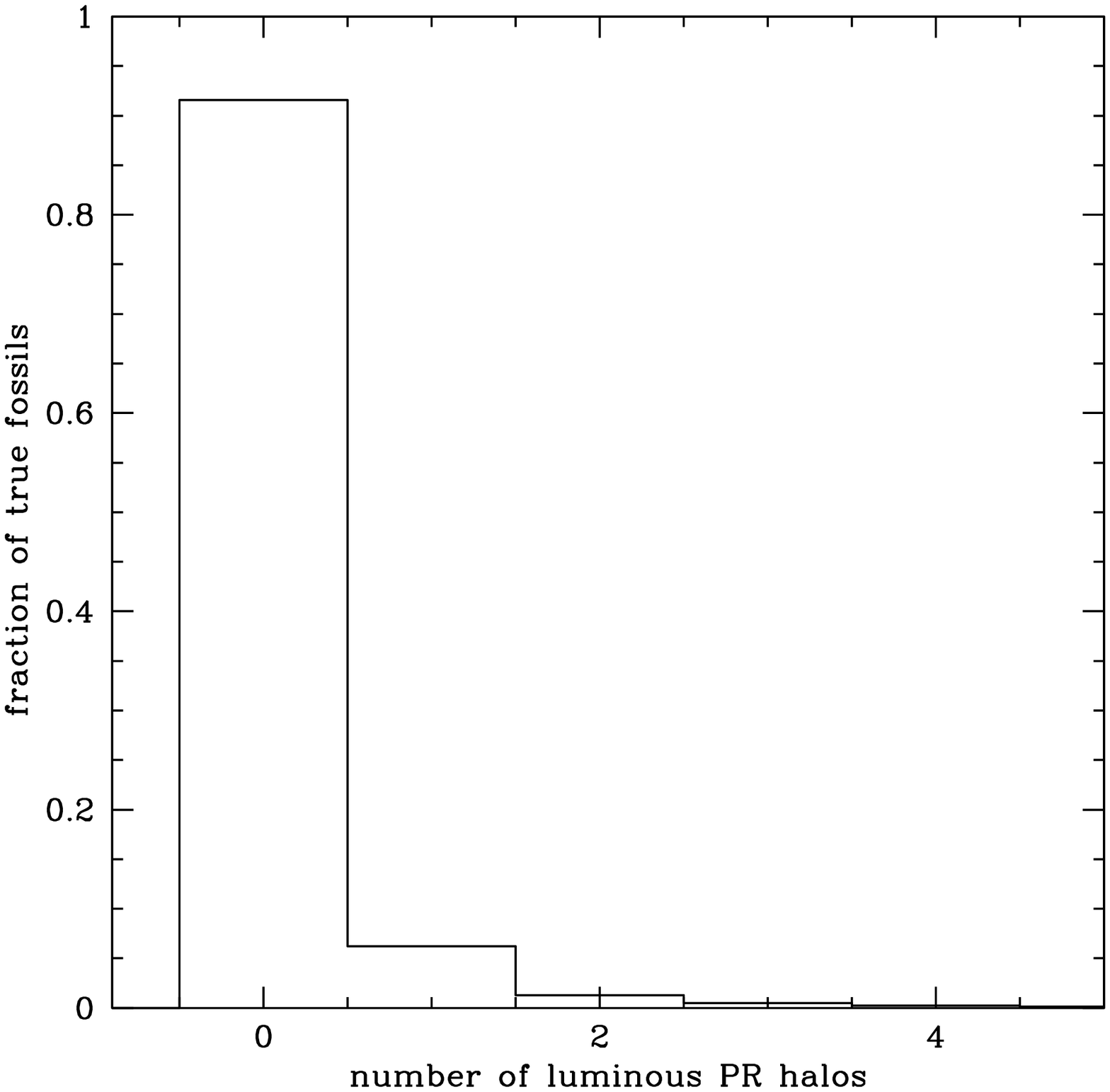}{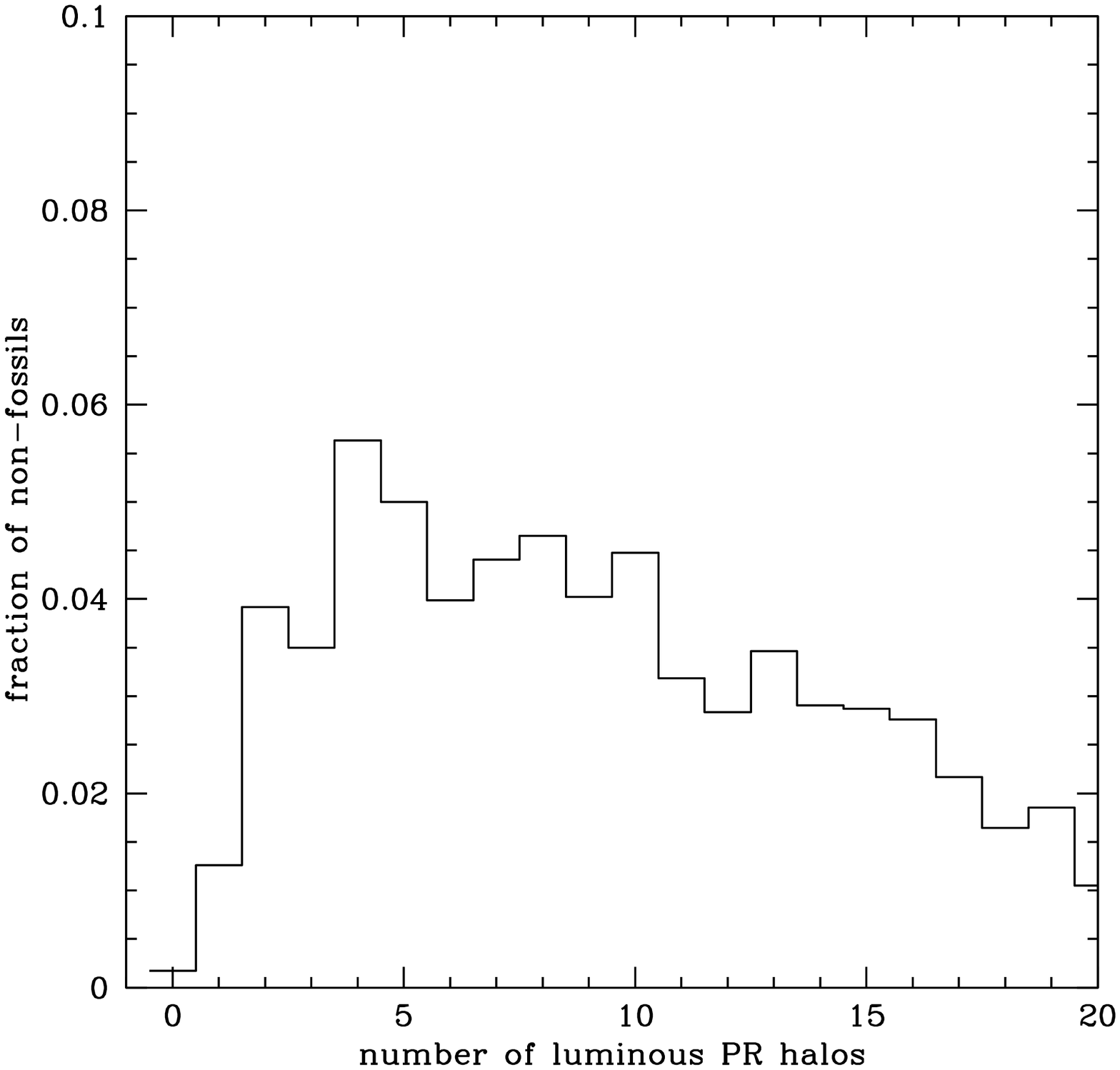}
\caption{({\it Left}). Histogram of the fraction of luminous true
  fossils with a given number of luminous pre-reionization halos,
  $N_{lum}$. $N_{lum}$ is a proxy for the number of significant
  mergers the system has undergone. ({\it{Right}}). Histogram of the
  fraction of non-fossils with a given $N_{lum}$. Note, that unlike
  the $N_{lum}$ histograms for the true fossils and polluted fossils
  the peak is not at $N_{lum} = 0$, but shifted to $N_{lum} \sim
  5$. Note also, that the vertical scale is 0.1 instead of 1.0.}
\label{Nlum.hist}
\end{figure*}

\begin{figure*}
\plottwo{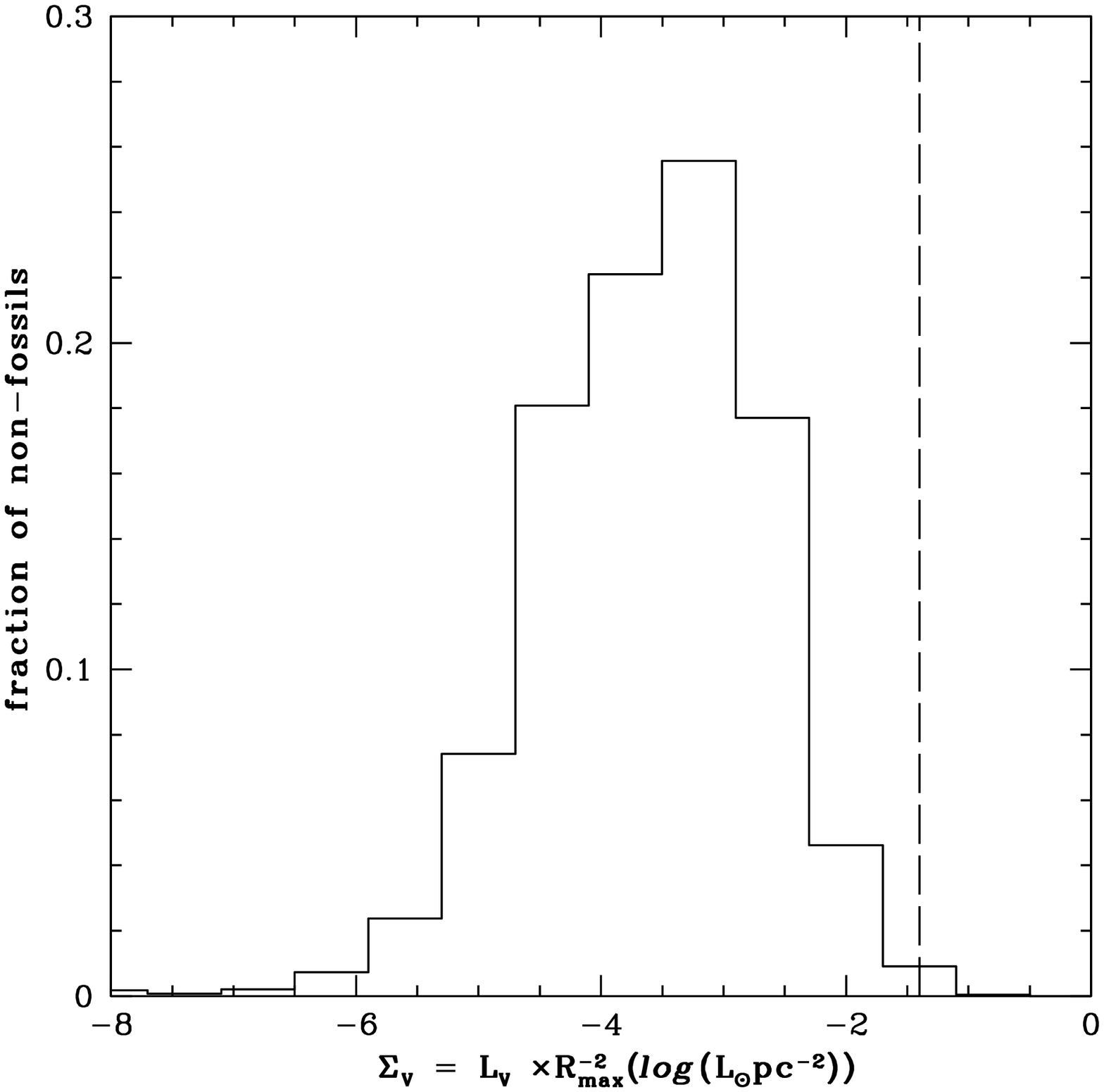}{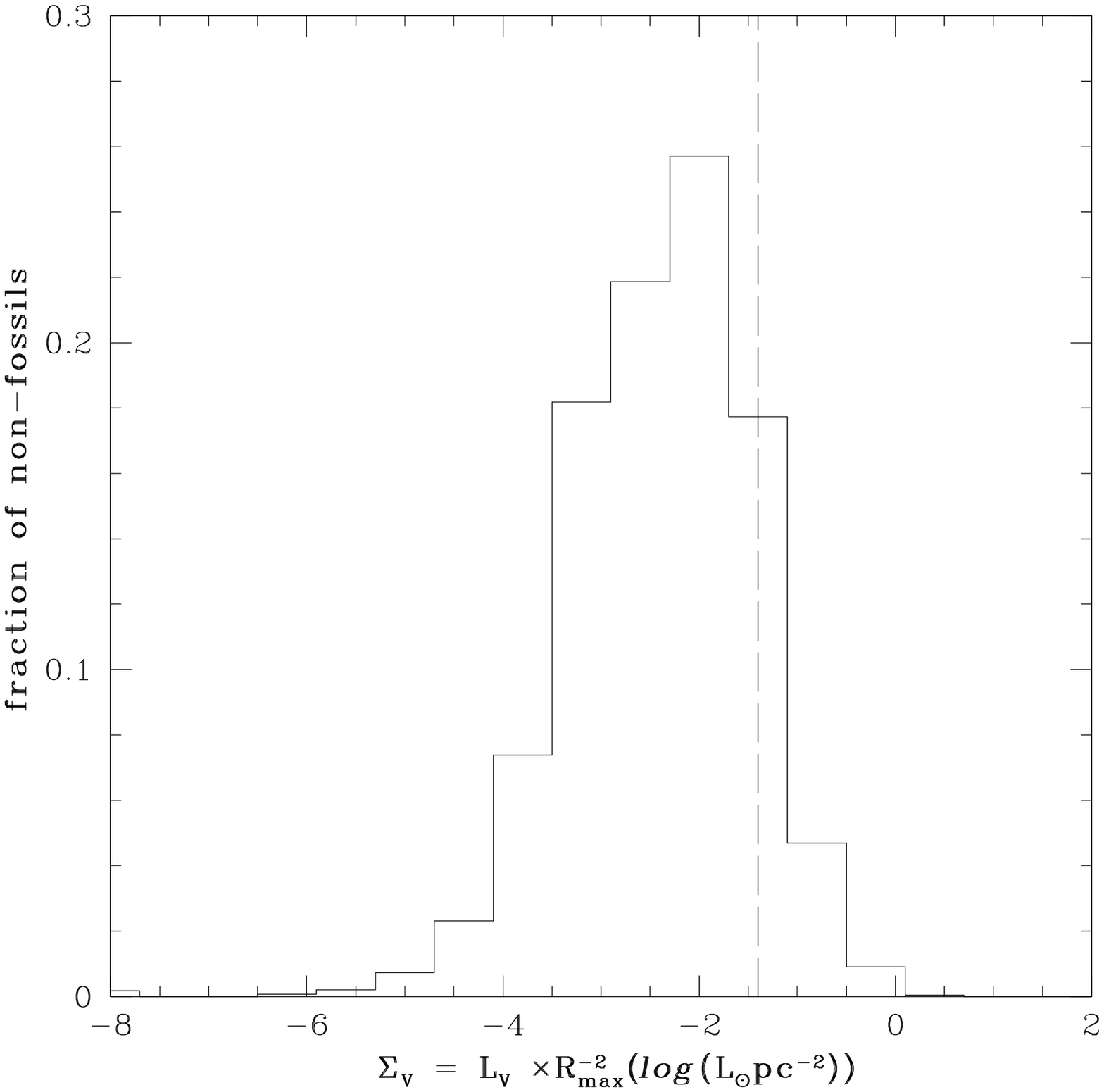}
\caption{({\it{Left}}) : Histogram of the fraction of non-fossils with a given V-band
  surface brightness, $\Sigma_V$, for the primordial population. We
  assume that the primordial stars have been puffed up by interactions
  until they fill the full extent of the dark matter halo, giving us
  $\Sigma_V = L_V \times R_{max}^{-2}$, where $R_{max}$ is the radius
  at which $v(r)=v_{max}$. The dashed vertical line is the surface
  brightness limit of the SDSS from \cite{Koposovetal:07}. Only
  $\sim1\%$ of the non-fossils are to the right of the dashed line,
  with expanded primordial populations detectable by the SDSS. ({\it{Right}}) : Histogram of the fraction of non-fossils with a given V-band
  surface brightness, $\Sigma_V$, for the primordial population. Here we assume that the ghost halo is more concentrated, extending only to $0.25\times R_{max}$.}
\label{SigmaV.nf}
\end{figure*}

\begin{figure*}
\centering
\plottwo{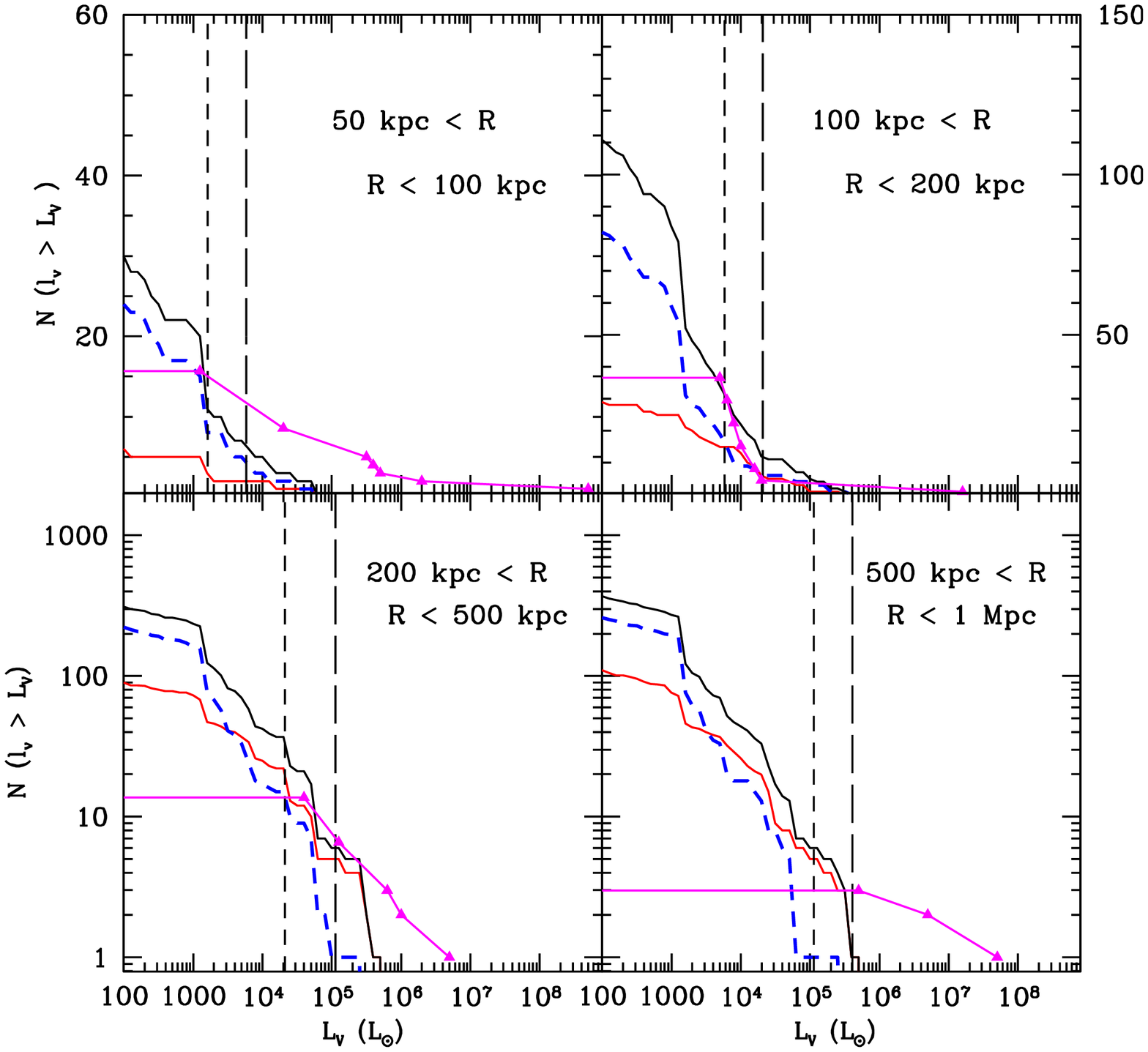}{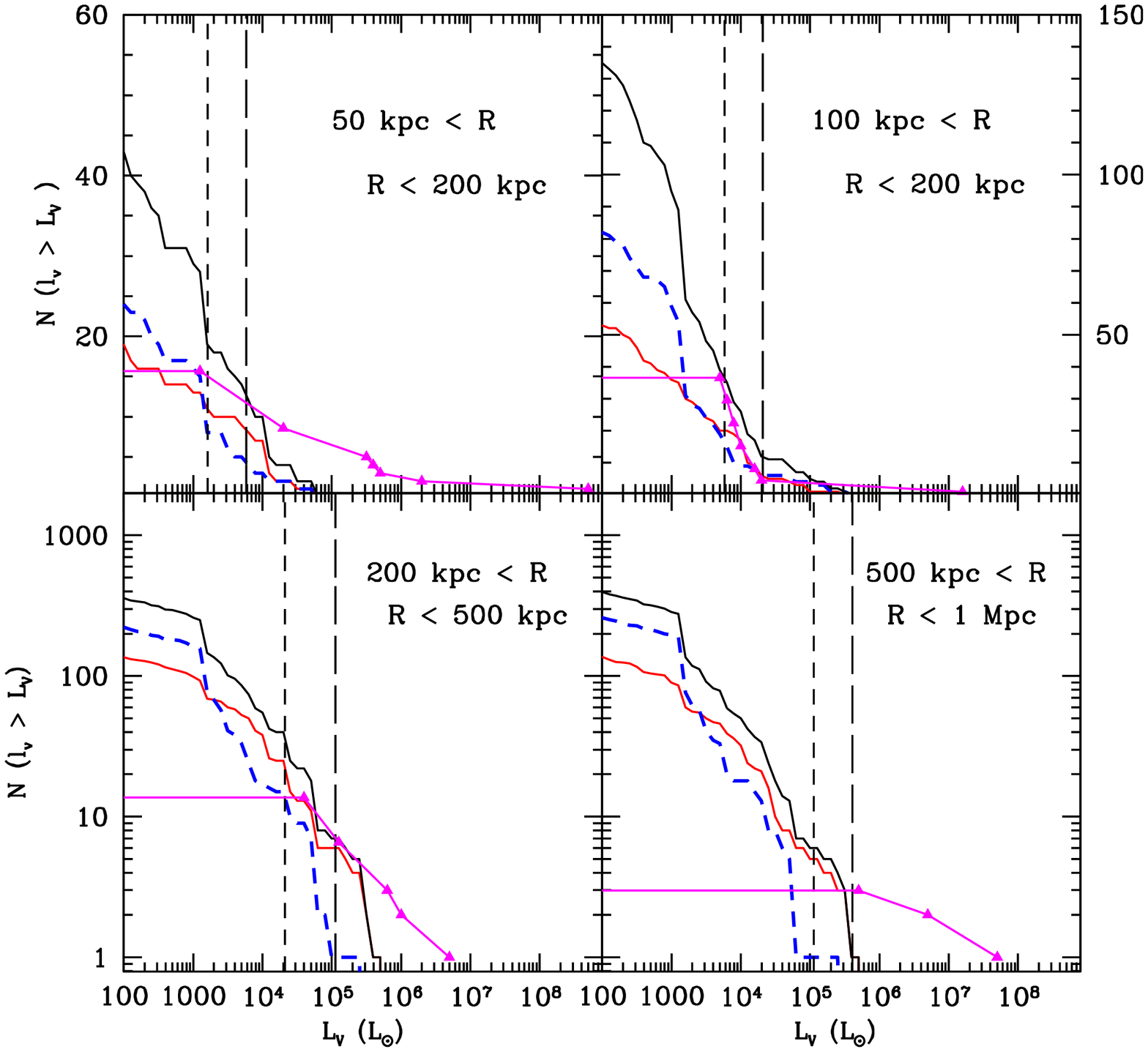}
\caption{{\it{Left}} : Cumulative luminosity functions of MW.3 from Run D (colored curves) with the total observed population (magenta triangles). All the symbols and lines mean the same as in Figure~\ref{LF.pristine}. In this figure we assume that all the non-fossils are below the detection limits or have lost their entire stellar populations due to a combination of heating due to major mergers and tidal interactions. {\it{Right}} : Cumulative luminosity functions of MW.3 from Run D (colored curves) with observations (magenta triangles). All the symbols and lines mean the same as in Figure~\ref{LF.pristine}. Unlike Figure~\ref{LF.nonf}, here we allow the non-fossils to retain $0.1\%$ of their stellar populations.}
\label{LF.nonf}
\end{figure*}

As discussed in Paper~I, our N-body method does not allow us to
determine the dynamics of the stars in halos that undergo mergers, or
the degree to which those stars are tidally stripped. However, we have
used analytic relationships to estimate the importance of dynamical
heating of the stars when $z=0$ fossils (about $20\%$) are produced by
mergers of more than one pre-reionization dwarf (Paper I) We refer to such interactions as galaxy mergers.  Here, we
focus on those dynamical processes in non-fossils which result in the
dispersion of the primordial stellar populations of the brightest
satellites, the net effect of which is to either make the non-fossil
populations invisible to current surveys by reducing their surface
brightnesses below the SDSS detection limits or preferentially
stripping them during interactions with more massive halos. The former
mechanism would be relevant to non-fossils at $R \simgt 500$~kpc where
tidal forces are negligible.

The number of pre-reionization halos in a $z=0$ dwarf
increases with mass. In this section, we explore the role of mergers
to rend invisible, or strip, the primordial populations of stars in the
the more massive dwarfs (non-fossils). Unlike in the previous
sections, here we differentiate between the non-fossils and polluted
fossils in our simulations. We remind the reader, that though both
populations have $v_{max}$ which were large enough for them to accrete
gas from the IGM in the past, only the non-fossils are at or above that threshold
at $z=0$.

When a system undergoes a galaxy merger, kinetic energy from the
collision is imparted to the stars. Immediately after the collision,
the new system will be in its most diffuse state. We define a galaxy
merger as the interaction of two or more pre-reionization
halos, both containing a primordial stellar population. Although there is significant scatter in the luminosities of
minihalos of the same mass, in general, the luminous minihalos are
more massive than those which are dark. An interaction between two
luminous minihalos is therefore more significant. We use $N_{lum}$,
the number of luminous pre-reionization halos within at $z=0$ halo, as
a proxy for the number of galaxy mergers. If there are multiple
galaxy mergers in a short amount of time, these stars will be
susceptible to stripping. In addition, recent work on increasing the
extent of the stellar population in bright ellipticals from $z=2$ to $z=0$ suggests that many
minor interactions over several Gyrs can increase the size of the
galaxy by a factor of 2-5 without significantly increasing mass
\citep{Naabetal:09}.  Roughly, the larger the number of significant
interactions, the greater the spatial extent of the pre-reionization
population and more likely the halo will have lost a significant
fraction of its primordial population to dynamical heating.

For isolated halos, with large $N_{lum}$, the radius of the primordial
population increases, possibly until it fills the spatial extent of
the dark matter halo. Such an extended system would have extremely low
surface brightness and would be susceptible to tidal stripping.  We next look
at which of our three subhalo populations has a significant number
of members with $N_{lum}>3$. The fractions of the true fossil, and
non-fossils populations with a given $N_{lum}$ are shown in the left
and right panels of Figure~\ref{Nlum.hist}. Neither of the fossil populations has a significant fraction of subhalos with $N_{lum}>3$ with
the fractions at $\sim1\%$ and $\sim10\%$ respectively. We therefore
assume, that, while a few of our true and polluted fossils may have
had their primordial populations diffused by mergers, the vast
majority remain dynamically cold.

The non-fossils show the opposite trend. The right panel of
Figure~\ref{Nlum.hist} shows that $<10\%$ of the non-fossils have
$N_{lum}<3$ and the distribution peaks at $N_{lum} \sim 5$. A
population of non-fossils would be much more likely to have a
primordial population dispersed by multiple major interactions than
their fossil counterparts.

The non-fossil populations (see Figure~\ref{LF.pristine}) could have
either lost their stellar populations, or had their primordial
populations increase in size to the point where they are undetectable
by the SDSS. Near the Milky Way, we assume they lost their entire
primordial stellar population to tidal interactions. After falling
into the Milky Way halo, the non-fossils were unable to form a
significant younger stellar population.

At larger radii, the non-fossils are less likely to have their
primordial populations stripped. However, as the stars expand to fill
the spatial extent of the dark matter halo, the non-fossils end up
with a primordial population with $\Sigma_V \sim L_V \times R_p^{-2}$,
where $L_V$ is the luminosity of the primordial population and $R_p$
is the radius of the primordial population. Figure~\ref{SigmaV.nf}
shows the fraction of the non-fossils with a primordial population
with surface brightness, $\Sigma_V$, for $R_p = R_{max}$ and $R_p =
0.25 \times R_{max}$, where $R_{max}$ is the radius of the maximum
circular velocity. We find that for $R_p = R_{max}$ about $1\%$ of
non-fossils would have an extended primordial halo above the SDSS
detection limits (to the right of the dashed line), and, when $R_p$ is
decreased to $0.25 \times R_{max}$, the detectable fraction only rises
to $~20\%$. If these non-fossils formed few or no stars after
reionization the majority would be undetectable by SDSS. The
non-fossils which did form stars after reionization today coulds be dIrr or one of the few isolated dSphs or dSphs/Irrs: \eg,
Cetus, Tucana, Antlia. In our scenario, they would be surrounded by
``ghost halos'' of primordial stars $\sim 12$~Gyr old with $[Fe/H]
\simlt -2$. But does the dispersal of the primordial population into ``ghost halos'' solve the ``bright satellite problem''?

We quantitatively approximate this for our simulations by using a
circular velocity cut. We look at the primordial luminosity functions
as if all the ghost halos are either stripped or below SDSS detection
limits. Practically, we set the luminosities to zero for all the
non-fossils. We look to see if this cut solves the bright satellite
problem while preserving the fossils better than lowering the star
formation efficiency. We find it to be a good solution to the bright
satellite problem.

Since setting the non-fossils luminosities to zero is able to decrease
the number of luminous satellites, we look at the luminosity function
it produces in more detail. All the curves in Figure~\ref{LF.nonf} are
the same as in Figure~\ref{LF.pristine}, excepting the red curve for
the non-fossils that now represents only the polluted fossils.

We now look at each distance bin to see what turning off the
non-fossil population has done to our various arguments.  For
$50$~kpc~$<R<100$~kpc, the necessity of a primordial dwarf population
is even stronger when we only consider the fossil and polluted fossil
populations. There are only $\sim 7$ polluted fossils within
$100$~kpc, less than one-fifth of what is required to account for the
$\sim 50$ observed satellites.  For $L_V > 10^4 L_\odot$, the
luminosity function now sits below observations. This gives the
remaining star forming halos room to form additional stars without
overproducing subhalos with $L_V>10^4 L_\odot$. The total number of subhalos
within $100$~kpc decreasing to 35 is consistent with
observations. First, our MW.3 is on the low end of the mass range for
the Milky Way for both observational estimates and
simulations. Second, as seen in Figure~\ref{LF.MF.part}, there are more
than enough stripped down fossils which formed in halos with $M >10^7
M_\odot$ and $L_V >10^5 L_\odot$ to fill in the deficit. Since these
objects would lack a cloud of tracer particles, they are marked as unbound by the halo finder AHF and
would be included in the luminosity and mass of the host halo and not in any luminosity function of the satellites.

The complete invisibility of the non-fossils is not quite as
successful for $100$~kpc$<R<200$~kpc as we are still slightly
overproducing the number of $L_V > 10^5$~L$_\odot$ satellites compared
to observations. However, we are better able to reproduce the sudden
steepening in the observed luminosity function in Figure~\ref{LF.nonf}
than with any of our other suppression mechanisms
(Figures~\ref{LF.ml500}, ~\ref{LF.high10}, and
especially~\ref{LF.sup}). This feature may be unique to the Milky Way
so we are not unduly concerned with matching it. In addition, if the
non-fossils are dark, our argument for the existence of primordial
fossils becomes straightforward. There are only $\sim 30$ polluted
fossils in this distance bin, only $75\%$ of the $\sim 40$ observed
galaxies, with any dwarf with $L_V<10^4$~L$_\odot$ difficult, if not
impossible, to detect with current surveys.

Beyond the MW.3 virial radius ($R\sim200$~kpc), turning the
non-fossils dark easily places the primordial luminosity function into
agreement with observations. This allows for the formation of
post-reionization populations of stars in the polluted fossils and
non-fossils. We remind the reader that the $z=0$ halos at these radii
would likely be on first approach to the Milky Way system and more
likely to accrete and retain gas at later times. The diffuse
primordial population in these distant non-fossils is an observational
test of star formation in pre-reionization dwarfs and the existence of
pre-reionization fossils.

\section{Three Observational Tests and Model Predictions}
\label{Tests}

Although the existence of pre-reionization fossils seems likely,
observations do not unequivocally demonstrate their existence due to
the large uncertainties in estimating the number of yet undiscovered
ultra-faint dwarfs \citep{Tollerudetal:08}.  In this section, we
summarize three observational tests for the existence of fossils
of the first galaxies that we propose based on the results in
Paper~I and the present work. The first test of our model is
especially interesting as it can be performed using HST observations
and does not require waiting for future all sky surveys deeper than
SDSS, like PanStar or LSST, to be online.

\begin{enumerate}

\item{\it{``Ghost halos'' around dwarfs on the outskirts of the Local Group}}

The primordial stellar populations in minihalos that formed before reionization should produce diffuse ``ghost halos'' of primordial stars around isolated dwarfs. We have shown that the total luminosity of the ``ghost halos'' is comparable to the one of classical dwarfs, but the surface brightness of the stars is well below the SDSS detection limits. Contrary to the difficulties of finding ultra-faints, we know where these diffuse stars are and we can plan deep observations to detect them. The diffuse primordial stellar populations around non-fossils should not be tidally stripped in dwarfs with galactocentric distance $>1$~Mpc from the Milky Way. We do not know within which distance tidal stripping would become important, but due to their large half light radii, they certainly are the stellar population that would be stripped first. ``Ghost halos'' can be best detected by resolving their individual main sequence stars around isolated dIrrs or dSphs before using spectra to determine their metallicities and dynamics. Unlike younger stars dispersed from the central galaxy, the primordial ghost halo would have a $[Fe/H]<-2.5$, and we are currently running simulations to determine their dynamics.

Recent HST observations of M31 have resolved the main sequence using ACS \citep{Brownetal:06,Brownetal:08,Brownetal:09}. A low-luminosity dwarf at $\sim 800$~kpc on the other side of the Milky Way from M31 would be an excellent candidate for the ghost halo search, and detection of its main-sequence primordial stars would be within the reach of HST. At $800$~kpc, the field of view of WFC3 (162'') is $\sim 630$~pc, at 1 Mpc, $\sim785$~pc, and at 2 Mpc, $\sim 1.6$~kpc. The ghost halos are $>1$~kpc in radius, often up to a few tens of kpc, therefore, WFC3 would not be able to image the entire dwarf with its ghost halo. However, aimed at the outskirts of a likely ghost halo host, it could look for signs of a primordial halo in the color magnitude diagram and radial surface brightness distribution. We would be able to resolve the individual stars in the ghost halos at $\sim 1$~Mpc. The ghost halos have surface densities of stars of 0.001 and 1~star~pc$^{-2}$ depending on the extent of the ghost halos and the slope of the IMF at low masses. Assuming a stellar density of $1$~star~pc$^{-2}$, the angular distance between each star is $\sim0.26''$, larger than the WFC3 resolution of $0.04''$ per pixel. Since the stellar population can be resolved, determining the details of a ghost halo population is a matter of taking deep enough exposure to detect the main sequence stars. Red giant branch stars, while brighter and easier to detect, have a density three orders of magnitude lower than the main sequence.

Deep observations work well for detecting the ghost halos when we already know where they are. These primordial populations surround dwarf galaxies that have undergone significant star formation since reionization and may have detectable gas and active star formation today. However, as discussed in \S~\ref{SEC.gh}, in order to reproduce the observed satellite distribution, only a fraction of the ghost halos can have formed a significant younger stellar populations. We remind the reader that our definition of a fossil versus a non-fossil in the simulations assumes a constant $v_{filter}=20$~km~s$^{-1}$. Assuming a larger filtering velocity produces a smaller number of non-fossils, some of the halos we define as non-fossils, and containing ghost halos, may have been unable to accrete gas and form stars after reionization due to additional heating of the IGM. This population of ``dead'' ghost halos would have only an extremely diffuse, primordial population. Without \HI~or more concentrated, younger stars, the best chance for detecting these ghost halos would be large scale surveys. Figure~\ref{SigmaV.nf} shows that the majority of the ghost halos are beyond the reach of the SDSS, but what about upcoming, deeper surveys such as PanSTARRS?

Diffuse stellar systems like the ultra-faint dwarfs and the ghost halos are found in surveys by looking for overdensities of stars relative to the background. In many cases, by detecting stars at the tip of the red giant branch (RGB). In low luminosity systems the detection of the RGB depends on two factors, the distance to the halo and the population of the RGB. Low luminosity systems, like the ultra-faint dwarfs, can have as few as a thousand stars, and therefore a sparsely populated, and difficult to detect, RGB. For example, with SDSS (magnitude limit r = 22.5) Hercules ($1.1\times 10^4 L_\odot$) could only be detected to $300$~kpc, while the more luminous CVn I ($2.3\times 10^5 L_\odot$) would be seen at a Mpc from the Milky Way \citep{Koposovetal:07}. PanSTARRS (magnitude limit r=24) will reach 1.5 magnitudes deeper that SDSS, detecting the same RGB twice as far.  However, the primordial populations in the ghost halos are {\it{extremely}} diffuse, and it is unclear that the overdensity of their RGB stars would be high enough to be detected against foreground M dwarfs and distant galaxies. In short, while PanSTARRS is expected to detect new ultra-faint dwarfs, it may not the best tool for finding ghost halos.

If some of the discovered ultra-faint dwarfs are fossils, then ``ghost halos'' should exist. Vice versa, the detection of ``ghost halos,'' regardless of the method, can be used to constrain the star formation rates before reionization and would imply the existence of fossils, although these fossils may not have yet been discovered due to their low surface brightnesses.

\item{\it{Population of yet undetected ultra-faint dwarfs}} 

  In BR09 and Paper~I we have discussed in detail the existence and
  the properties of a yet undetected population of fossils with
  $L_V<10^4$~L$_\odot$ and surface brightness $\Sigma_V<
  10^{-1.4}$~L$_\odot$~pc$^{-2}$.  This population of ultra-faints
  should be accessible to future all sky survery such as PanStars and
  LSST. If the undetected dwarfs are fossils, they should have stellar
  velocity dispersions equivalent to those of the ultra-faints with
  corresponding mass to light ratios of $10^4 M_\odot/L_\odot$ or
  higher. The typical $[Fe/H]$ of the ``stealth'' fossils should be
  $<-2.5$. We direct the reader to Paper I for a detailed
  justification of these predictions. 

\item{\it{Dark and ultra-faint gas rich dwarfs in the voids}}

  According to the model proposed in \cite{Ricotti:09}, a subset of
  minihalos in the voids may have been able to condense gas from the
  IGM after Helium~II reionization (at $z\sim3$). However, they would
  not form stars unless their gas reached a sufficient density. These
  minihalos may or may not have formed stars before reionization, and
  any stellar populations they did have would be below the detection
  limits of both current and future surveys. The HI in these objects
  could be detected by blind 21 cm surveys. Recently, ALFALFA and
  GALFA surveys have reported the discovery of several small and
  compact clouds of neutral hydrogen, some of which may represent a
  population of pre-reionization minihalos. Some of these clouds could
  be ``dark galaxies'' and represent the smallest detectable halos
  around the Milky Way, others may be ultra-faint dwarfs in the voids.
  The location of unassociated HI detections could then guide optical
  surveys to these primordial fossils in a focused deep search for
  their ancient populations.
\end{enumerate}

\section{Summary and Discussion}
\label{Disc}

Through this work and Paper I, we have used results of simulations to
study the origin of observed Milky Way and M31 satellites and understand
whether they are compatible with models of star formation before
reionization. In the primordial model, a subset of the Milky Way
satellites formed with their current properties with minimal
modifications by tidal stripping. These low luminosity satellites
formed the majority ($>70\%$) of their stars in minihalos before
reionization. Our simulated true fossils produce an excellent
agreement in properties and distribution with the observed lowest
luminosity Milky Way satellites. In BR09 and Paper I, we showed that a
subset of the ultra-faints, all with $L_V < 10^5$~L$_\odot$, have
half-light radii, surface brightnesses, mass-to-light ratios, velocity
dispersions, metallicities and metallicity dispersions consistent with
the expected stellar properties of the true fossils. In the present
paper, we have compared the galactocentric radial distributions and
primordial cumulative luminosity functions of simulated fossils to
observations of Milky Way satellites. When we compare the observed and
simulated distributions, we find them to be in agreement with each other
for $L_V< 10^5$~L$_\odot$. In addition, a large population of primordial fossils have surface
brightness below the detection limits of current surveys (\eg, SDSS).
The following list summarizes the main results of this second paper of
the series.
\begin{itemize}
\item We are able to reproduce the distribution of the ultra-faints
  with our simulated primordial fossils. We find no missing satellites
  at the lower end of the mass and luminosity functions, but our model
  predicts $\sim 150$ additional Milky Way satellites detectable by upcoming surveys (PanStars, LSST). In addition to the undiscovered ``missing'' satellites at distances too great (> 50-100 kpc) to be detected by SDSS or in areas on the sky not in the SDSS footprint, there is also a new population of ultra-faints with mega-faint surface brightness hidden below the SDSS detection limits at all distances from the Milky Way.

\item At all radii, we find an overabundance of simulated bright ($L_V > 10^4 L\odot$) satellites with respect to observations, which, even with only their primordial luminosities, would be easily detected by current surveys if we assume they have the same half-light radii as the other known ultra-faint dwarfs. Given the agreement between the stellar properties and distributions of the ultra-faints and those of our fossil dwarfs, we cannot account for the excess bright satellites by imposing a blanket suppression of star formation below a given mass.

\item Lower $H_2$ formation rates and subsequent lower minihalo star
  formation rates in the voids are not able to bring the number of
  bright satellites into agreement with observations.

\item Effectively lowering the star formation efficiency can fix the
  bright satellite problem if we assume only pre-reionization halos
  with $M < 7\times10^6$~M$_\odot$ had SFR dominated by local,
  stochastic feedback. However, not only is this contrary to current
  understanding of star formation in minihalos, but the fossil population
  this ``solution'' produces cannot reproduce the distribution of the
  ultra-faint population.

\item We bring the number of bright satellites into agreement with
  observations, while leaving the fossil population untouched, by
  assuming the primordial stellar populations of our non-fossils (with
  maximum circular velocities, $v_{max}(z=0) > v_{filt}$) become
  extremely diffuse via kinetic energy from galaxy mergers. The
  existence of ``ghost halos'' of primordial stars is a new powerful
  observational prediction of our model that can be straight forwardly tested
  using HST observations of isolated dwarfs around the Local Group.

\end{itemize}

One of the key predictions of the primordial model is a total number
of satellites for the Milky Way between $200-300$, only a maximum of
100 of which are non-fossils (here we assumed
$v_{filt}=20$~km~s$^{-1}$).  The number of Milky Way satellites which
are not fossils provides an important test for star formation in
minihalos at high redshift. If, after PanSTARRS and LSST are online,
the number of ultra-faint Milky Way satellites remains $<100$, we have
a strong constraint on star formation in pre-reionization
dwarfs. Either no pre-reionization fossils survived near the Milky
Way, or almost none of the halos with masses at formation $M<10^8
M_\odot$ formed stars. However, if the satellite count rises to
$>100$, some of the dimmest Milky Way dwarfs must be fossils of
reionization. Using details of the stellar populations and their
distributions, observations of these fossils can constrain models of
star formation at high redshift.

A caveat to this picture is that the number of non-fossils is highly
sensitive to the choice of the filtering velocity. When we raise the
filtering velocity to $30$~km~s$^{-1}$, the number of non-fossils
drops by a third to $60\pm8$ from the $90\pm10$ for
$v_{filt}=20$~km~s$^{-1}$. The choice of $20$~km~s$^{-1}$ assumes a
constant IGM density with $T_{IGM}=10^4$~K throughout a minihalo's
evolution. In reality, the situation is not so simple. The gas near
$10^{12}$~M$_\odot$ halos and in the filaments between may be heated
to $\sim 10^5-10^6$~K by AGN feedback.
The higher temperatures of this local intergalactic medium may
correspond to $v_{filt}\simgt 40$~km~s$^{-1}$. In addition to the
higher filtering velocity, the higher density near a Milky Way mass
halo reduces the effective potential depth of the subhalos, increasing
the mass threshold for post-reionization gas accretion still
further. Simulations to determine the temperature and density of the
IGM near a Milky Way from reionization to the modern epoch are needed
to determine the $v_{filt}({\bf{x}},z)$, and whether these factors can
explain the existence of Milky Way and M31 dwarfs with the observed
properties of fossils, but luminosities above the $10^6$~L$_\odot$
threshold.

The observed distributions we compare to depend on how we correct for
the incomplete sky coverage of the SDSS. In this work, we have assumed
an isotropic satellite distribution at $R>50$~kpc. Under this
assumption, the SDSS completeness correction for the ultra-faints is
$3.54$. We briefly check if the agreement between the observed and
simulated distributions is dependent on the isotropic assumption.
Recent work \citep{Metzetal:07,Metzetal:09} has suggested that rather
than being isotropic, the Milky Way satellites are oriented in a plane
approximately perpendicular to the disk.  We approximate this
non-homogeneous satellite distribution by correcting for the SDSS sky
coverage by a factor of 2.0 instead of 3.54. The number of classical
fossils remains the same. The different correction does not change the
consistency of our simulated galactocentric distribution with
observations, though the lower correction factor suggests a higher
Milky Way mass. It also does not change the bright satellite problem,
in fact, the lower observational correction factor makes the
overabundance of simulated $L_V > 10^4$~L$_\odot$ dwarfs worse by
about a factor of two.

We have suggested two solutions which correct for the
overabundance of bright satellites while preserving at least a
fraction of the primordial fossil population. Each presents a
different picture when we consider it in the context of the voids. The
first, and less effective, solution calls for a low star formation
efficiency. In this picture, the $10^{10}$~M$_\odot$ halos visible in
current surveys will have their star formation dampened, however as we
move to $~10^7$~M$_\odot$ we enter the regime where stochastic
feedback effects dominate over the choice of $\epsilon_\ast$. Thus,
the voids would appear relatively empty, but only because we cannot
yet detect the less than $10^5$~L$_\odot$ fossil populations which
formed in the $10^7$~M$_\odot$ halos before reionization.
	
The dispersal of the non-fossils' primordial populations into ghost
halos is a more effective solution to the ``bright satellite problem''
within $1$~Mpc of the Milky Way, but leaves a conundrum in the
voids. Regardless of whether the primordial population would be
detectable, how do we keep the post-reionization star formation in
these non-fossils low enough to prevent this later star formation from
producing more $M_V > -16$ galaxies than are currently observed?

Any post-reionization star formation in the non-fossils results in a young population which would be (i) brighter and bluer, (ii) more concentrated, since enriched gas will cool faster and sink deeper into the gravitational potential, and (iii) possibly accompanied by an \HI reservoir. Any of these properties would make the post-reionization population easier to see, and the the non-fossil harder to hide. To suppress the post-reionization baryonic evolution in the non-fossils we examine our naive assumption that they all undergo significant baryonic evolution after reionization. 

The easiest way to suppress star formation in the lower mass non-fossils is to raise the filtering velocity. As has already been discussed in Chapter 4, any non-fossil embedded in the WHIM ($T\sim10^5$~K) would have a $v_{filter}\sim40$~km~s$^{-1}$. However, the WHIM does not exist until $z<1$ \citep{Smithetal:10}, leaving $\sim6$~Gyr after reionization when the non-fossils could have accreted gas and formed stars. Active galactic nuclei (AGN) eject enormous amounts of energy into their environs, heating the gas and raising the filtering velocity, but how far from the host galaxy the AGN is effective at suppressing star formation in dwarfs ,and for how long is unclear. In addition, we see a bright satellite problem around every large Local Volume galaxy. Is it reasonable to assume that, at some point in its evolution, every $L_\ast$ galaxy hosted an AGN? A final possibility is that reionization was extremely efficient at quenching star formation in $20-40$~km~s$^{-1}$ halos and the non-fossils were never able to build up enough gas from the post-reionization IGM to form additional stars.

In summary, while the bright satellite problem can be ``solved'' for the primordial population alone, we still need to account for the post-reionization evolution of the non-fossils. In order to maintain the agreement with observations, only $\sim10\%$ of the non-fossils can form significant stellar populations after reionization. Determining how and if the other $\sim90\%$ can be suppressed will tell us how much of a problem the bright satellite problem is. 

\acknowledgements

The simulations presented in this paper were carried out using computing clusters administered by the Center for Theory and Computation of the Department of Astronomy at the University of Maryland ("yorp"), and the Office of Information Technology at the University of Maryland ("hpcc"). This research was supported by NASA grants NNX07AH10G and NNX10AH10G. The authors thank the anonymous referee for constructive comments and feedback. Thanks from MSB and MR to Stacy McGaugh, Derek Richardson and Rosie Wyse for helpful conversations and comments. MSB would like to thank Susan Lamb for discussions of dwarf dynamics and Evan Kirby and Beth Willman for discussions of modeling and observations.

\bibliographystyle{./apj}
\bibliography{../../refs}

\begin{thebibliography}{}

\bibitem[\protect\citeauthoryear{{Abazajian} et~al.}{{Abazajian}
  et~al.}{2009}]{Abazajianetal:09}
{Abazajian}, K.~N., et~al. 2009, \apjs, 182, 543

\bibitem[\protect\citeauthoryear{{Ahn} et~al.}{{Ahn} et~al.}{2006}]{Ahnetal:06}
{Ahn}, K., {Shapiro}, P.~R., {Alvarez}, M.~A., {Iliev}, I.~T., {Martel}, H.,
  \& {Ryu}, D. 2006, New Astronomy Review, 50, 179

\bibitem[\protect\citeauthoryear{{Bailin} et~al.}{{Bailin}
  et~al.}{2008}]{Bailinetal:08}
{Bailin}, J., {Power}, C., {Norberg}, P., {Zaritsky}, D.,  \& {Gibson}, B.~K.
  2008, \mnras, 390, 1133

\bibitem[\protect\citeauthoryear{{Barkana} \& {Loeb}}{{Barkana} \&
  {Loeb}}{2004}]{BarkanaL:04}
{Barkana}, R.,  \& {Loeb}, A. 2004, \apj, 609, 474

\bibitem[\protect\citeauthoryear{{Belokurov} et~al.}{{Belokurov}
  et~al.}{2007}]{Belokurovetal:07}
{Belokurov}, V., et~al. 2007, \apj, 654, 897

\bibitem[\protect\citeauthoryear{{Belokurov} et~al.}{{Belokurov}
  et~al.}{2006}]{Belokurovetal:06a}
{Belokurov}, V., et~al. 2006, \apjl, 647, L111

\bibitem[\protect\citeauthoryear{{Bovill} \& {Ricotti}}{{Bovill} \&
  {Ricotti}}{2009}]{BovillRicotti:09}
{Bovill}, M.~S.,  \& {Ricotti}, M. 2009, \apj, 693, 1859

\bibitem[\protect\citeauthoryear{{Bozek}, {Wyse}, \& {Gilmore}}{{Bozek}
  et~al.}{2011}]{Bozeketal:11AAS}
{Bozek}, B., {Wyse}, R.~F.~G.,  \& {Gilmore}, G.~F. 2011, in Bulletin of the
  American Astronomical Society, Vol.~43, American Astronomical Society Meeting
  Abstracts \#217, 147.06

\bibitem[\protect\citeauthoryear{{Brown} et~al.}{{Brown}
  et~al.}{2008}]{Brownetal:08}
{Brown}, T.~M., et~al. 2008, \apjl, 685, L121

\bibitem[\protect\citeauthoryear{{Brown} et~al.}{{Brown}
  et~al.}{2009}]{Brownetal:09}
{Brown}, T.~M., et~al. 2009, \apjs, 184, 152

\bibitem[\protect\citeauthoryear{{Brown} et~al.}{{Brown}
  et~al.}{2006}]{Brownetal:06}
{Brown}, T.~M., {Smith}, E., {Ferguson}, H.~C., {Rich}, R.~M., {Guhathakurta},
  P., {Renzini}, A., {Sweigart}, A.~V.,  \& {Kimble}, R.~A. 2006, \apj, 652,
  323

\bibitem[\protect\citeauthoryear{{Bullock}, {Kravtsov}, \&
  {Weinberg}}{{Bullock} et~al.}{2001}]{Bullocketal:01}
{Bullock}, J.~S., {Kravtsov}, A.~V.,  \& {Weinberg}, D.~H. 2001, \apj, 548, 33

\bibitem[\protect\citeauthoryear{{Bullock} et~al.}{{Bullock}
  et~al.}{2010}]{Bullocketal:10}
{Bullock}, J.~S., {Stewart}, K.~R., {Kaplinghat}, M., {Tollerud}, E.~J.,  \&
  {Wolf}, J. 2010, \apj, 717, 1043

\bibitem[\protect\citeauthoryear{{Ciardi}, {Ferrara}, \& {Abel}}{{Ciardi}
  et~al.}{2000}]{Ciardietal:00}
{Ciardi}, B., {Ferrara}, A.,  \& {Abel}, T. 2000, \apj, 533, 594

\bibitem[\protect\citeauthoryear{{Cioni} \& {Habing}}{{Cioni} \&
  {Habing}}{2005}]{CioniHabing:05}
{Cioni}, M.-R.~L.,  \& {Habing}, H.~J. 2005, \aap, 442, 165

\bibitem[\protect\citeauthoryear{{de Jong} et~al.}{{de Jong}
  et~al.}{2008}]{deJongetal:08}
{de Jong}, J.~T.~A., et~al. 2008, \apj, 680, 1112

\bibitem[\protect\citeauthoryear{{Efstathiou}}{{Efstathiou}}{1992}]{Efstathiou:92}
{Efstathiou}, G. 1992, \mnras, 256, 43P

\bibitem[\protect\citeauthoryear{{Frebel} et~al.}{{Frebel}
  et~al.}{2010}]{Frebeletal:10}
{Frebel}, A., {Simon}, J.~D., {Geha}, M.,  \& {Willman}, B. 2010, \apj, 708,
  560

\bibitem[\protect\citeauthoryear{{Geha} et~al.}{{Geha}
  et~al.}{2009}]{Gehaetal:09}
{Geha}, M., {Willman}, B., {Simon}, J.~D., {Strigari}, L.~E., {Kirby}, E.~N.,
  {Law}, D.~R.,  \& {Strader}, J. 2009, \apj, 692, 1464

\bibitem[\protect\citeauthoryear{{Haiman}, {Abel}, \& {Rees}}{{Haiman}
  et~al.}{2000}]{Haimanetal:00}
{Haiman}, Z., {Abel}, T.,  \& {Rees}, M.~J. 2000, \apj, 534, 11

\bibitem[\protect\citeauthoryear{{Irwin} et~al.}{{Irwin}
  et~al.}{2007}]{Irwinetal:07}
{Irwin}, M.~J., et~al. 2007, \apjl, 656, L13

\bibitem[\protect\citeauthoryear{{Karachentsev} et~al.}{{Karachentsev}
  et~al.}{2006}]{Karachentsevetal:06}
{Karachentsev}, I.~D., et~al. 2006, \aj, 131, 1361

\bibitem[\protect\citeauthoryear{{Karachentsev} et~al.}{{Karachentsev}
  et~al.}{2004}]{Karachentsevetal:04}
{Karachentsev}, I.~D., {Karachentseva}, V.~E., {Huchtmeier}, W.~K.,  \&
  {Makarov}, D.~I. 2004, \aj, 127, 2031

\bibitem[\protect\citeauthoryear{{Klypin} et~al.}{{Klypin}
  et~al.}{1999}]{Klypinetal:99}
{Klypin}, A., {Kravtsov}, A.~V., {Valenzuela}, O.,  \& {Prada}, F. 1999, \apj,
  522, 82

\bibitem[\protect\citeauthoryear{{Knollmann} \& {Knebe}}{{Knollmann} \&
  {Knebe}}{2009}]{KnollmannKnebe:09}
{Knollmann}, S.~R.,  \& {Knebe}, A. 2009, \apjs, 182, 608

\bibitem[\protect\citeauthoryear{{Koposov} et~al.}{{Koposov}
  et~al.}{2008}]{Koposovetal:07}
{Koposov}, S., et~al. 2008, \apj, 686, 279

\bibitem[\protect\citeauthoryear{{Kravtsov}}{{Kravtsov}}{2010}]{Kravtsov:10}
{Kravtsov}, A. 2010, Advances in Astronomy, 2010

\bibitem[\protect\citeauthoryear{{Kroupa}, {Theis}, \& {Boily}}{{Kroupa}
  et~al.}{2005}]{Kroupaetal:05}
{Kroupa}, P., {Theis}, C.,  \& {Boily}, C.~M. 2005, \aap, 431, 517

\bibitem[\protect\citeauthoryear{{Macci{\`o}} et~al.}{{Macci{\`o}}
  et~al.}{2010}]{Maccioetal:10}
{Macci{\`o}}, A.~V., {Kang}, X., {Fontanot}, F., {Somerville}, R.~S.,
  {Koposov}, S.,  \& {Monaco}, P. 2010, \mnras, 402, 1995

\bibitem[\protect\citeauthoryear{{Machacek} et~al.}{{Machacek}
  et~al.}{2000}]{Machaceketal:00}
{Machacek}, M.~E., {Bryan}, G.~L., {Meiksin}, A., {Anninos}, P., {Thayer}, D.,
  {Norman}, M.,  \& {Zhang}, Y. 2000, \apj, 532, 118

\bibitem[\protect\citeauthoryear{{Maio} et~al.}{{Maio}
  et~al.}{2010}]{Maioetal:10}
{Maio}, U., {Ciardi}, B., {Dolag}, K., {Tornatore}, L.,  \& {Khochfar}, S.
  2010, \mnras, 905

\bibitem[\protect\citeauthoryear{{Martinez} et~al.}{{Martinez}
  et~al.}{2010}]{Martinezetal:10}
{Martinez}, G.~D., {Minor}, Q.~E., {Bullock}, J., {Kaplinghat}, M., {Simon},
  J.~D.,  \& {Geha}, M. 2010, ArXiv e-prints

\bibitem[\protect\citeauthoryear{{Mateo}}{{Mateo}}{1998}]{Mateo:98}
{Mateo}, M.~L. 1998, \araa, 36, 435

\bibitem[\protect\citeauthoryear{{McGaugh} et~al.}{{McGaugh}
  et~al.}{2010}]{McGaughetal:10}
{McGaugh}, S.~S., {Schombert}, J.~M., {de Blok}, W.~J.~G.,  \& {Zagursky},
  M.~J. 2010, \apjl, 708, L14

\bibitem[\protect\citeauthoryear{{McGaugh} \& {Wolf}}{{McGaugh} \&
  {Wolf}}{2010}]{McGaughW:10}
{McGaugh}, S.~S.,  \& {Wolf}, J. 2010, ArXiv e-prints

\bibitem[\protect\citeauthoryear{{Metz}, {Kroupa}, \& {Jerjen}}{{Metz}
  et~al.}{2007}]{Metzetal:07}
{Metz}, M., {Kroupa}, P.,  \& {Jerjen}, H. 2007, \mnras, 374, 1125

\bibitem[\protect\citeauthoryear{{Metz}, {Kroupa}, \& {Jerjen}}{{Metz}
  et~al.}{2009}]{Metzetal:09}
{Metz}, M., {Kroupa}, P.,  \& {Jerjen}, H. 2009, \mnras, 394, 2223

\bibitem[\protect\citeauthoryear{{Monelli} et~al.}{{Monelli}
  et~al.}{2010}]{Monellietal:10}
{Monelli}, M., et~al. 2010, \apj, 720, 1225

\bibitem[\protect\citeauthoryear{{Moore} et~al.}{{Moore}
  et~al.}{1999}]{Mooreetal:99}
{Moore}, B., {Ghigna}, S., {Governato}, F., {Lake}, G., {Quinn}, T., {Stadel},
  J.,  \& {Tozzi}, P. 1999, \apjl, 524, L19

\bibitem[\protect\citeauthoryear{{Naab}, {Johansson}, \& {Ostriker}}{{Naab}
  et~al.}{2009}]{Naabetal:09}
{Naab}, T., {Johansson}, P.~H.,  \& {Ostriker}, J.~P. 2009, \apjl, 699, L178

\bibitem[\protect\citeauthoryear{{Niederste-Ostholt}
  et~al.}{{Niederste-Ostholt} et~al.}{2009}]{Niederste-Ostholtetal:09}
{Niederste-Ostholt}, M., {Belokurov}, V., {Evans}, N.~W., {Gilmore}, G.,
  {Wyse}, R.~F.~G.,  \& {Norris}, J.~E. 2009, \mnras, 398, 1771

\bibitem[\protect\citeauthoryear{{Norris} et~al.}{{Norris}
  et~al.}{2010}]{Norrisetal:10}
{Norris}, J.~E., {Wyse}, R.~F.~G., {Gilmore}, G., {Yong}, D., {Frebel}, A.,
  {Wilkinson}, M.~I., {Belokurov}, V.,  \& {Zucker}, D.~B. 2010, ArXiv e-prints

\bibitem[\protect\citeauthoryear{{O'Shea} \& {Norman}}{{O'Shea} \&
  {Norman}}{2008}]{OSheaN:08}
{O'Shea}, B.~W.,  \& {Norman}, M.~L. 2008, \apj, 673, 14

\bibitem[\protect\citeauthoryear{{Peebles}}{{Peebles}}{2001}]{Peebles:01}
{Peebles}, P.~J.~E. 2001, \apj, 557, 495

\bibitem[\protect\citeauthoryear{{Ricotti}}{{Ricotti}}{2009}]{Ricotti:09}
{Ricotti}, M. 2009, \mnras, 392, L45

\bibitem[\protect\citeauthoryear{{Ricotti}}{{Ricotti}}{2010}]{Ricotti:10}
{Ricotti}, M. 2010, Advances in Astronomy, 2010

\bibitem[\protect\citeauthoryear{{Ricotti} \& {Gnedin}}{{Ricotti} \&
  {Gnedin}}{2005}]{RicottiGnedin:05}
{Ricotti}, M.,  \& {Gnedin}, N.~Y. 2005, \apj, 629, 259

\bibitem[\protect\citeauthoryear{{Ricotti}, {Gnedin}, \& {Shull}}{{Ricotti}
  et~al.}{2000}]{RicottiGnedinShull:00}
{Ricotti}, M., {Gnedin}, N.~Y.,  \& {Shull}, J.~M. 2000, \apj, 534, 41

\bibitem[\protect\citeauthoryear{{Ricotti}, {Gnedin}, \& {Shull}}{{Ricotti}
  et~al.}{2001}]{RicottiGnedinShull:01}
{Ricotti}, M., {Gnedin}, N.~Y.,  \& {Shull}, J.~M. 2001, \apj, 560, 580

\bibitem[\protect\citeauthoryear{{Ricotti}, {Gnedin}, \& {Shull}}{{Ricotti}
  et~al.}{2002a}]{RicottiGnedinShull:02a}
{Ricotti}, M., {Gnedin}, N.~Y.,  \& {Shull}, J.~M. 2002a, \apj, 575, 33

\bibitem[\protect\citeauthoryear{{Ricotti}, {Gnedin}, \& {Shull}}{{Ricotti}
  et~al.}{2002b}]{RicottiGnedinShull:02b}
{Ricotti}, M., {Gnedin}, N.~Y.,  \& {Shull}, J.~M. 2002b, \apj, 575, 49

\bibitem[\protect\citeauthoryear{{Ricotti}, {Gnedin}, \& {Shull}}{{Ricotti}
  et~al.}{2008}]{RicottiGnedinShull:08}
{Ricotti}, M., {Gnedin}, N.~Y.,  \& {Shull}, J.~M. 2008, \apj, 685, 21

\bibitem[\protect\citeauthoryear{{Ricotti} \& {Ostriker}}{{Ricotti} \&
  {Ostriker}}{2004}]{RicottiOstriker:04}
{Ricotti}, M.,  \& {Ostriker}, J.~P. 2004, \mnras, 352, 547

\bibitem[\protect\citeauthoryear{{Ricotti}, {Ostriker}, \& {Gnedin}}{{Ricotti}
  et~al.}{2005}]{RicottiOstrikerGnedin:05}
{Ricotti}, M., {Ostriker}, J.~P.,  \& {Gnedin}, N.~Y. 2005, \mnras, 357, 207

\bibitem[\protect\citeauthoryear{{Sand} et~al.}{{Sand}
  et~al.}{2009}]{Sandetal:09}
{Sand}, D.~J., {Olszewski}, E.~W., {Willman}, B., {Zaritsky}, D., {Seth}, A.,
  {Harris}, J., {Piatek}, S.,  \& {Saha}, A. 2009, \apj, 704, 898

\bibitem[\protect\citeauthoryear{{Simon} \& {Geha}}{{Simon} \&
  {Geha}}{2007}]{SimonGeha:07}
{Simon}, J.~D.,  \& {Geha}, M. 2007, \apj, 670, 313

\bibitem[\protect\citeauthoryear{{Simon} et~al.}{{Simon}
  et~al.}{2010}]{Simonetal:10}
{Simon}, J.~D., et~al. 2010, ArXiv e-prints

\bibitem[\protect\citeauthoryear{{Smith} et~al.}{{Smith}
  et~al.}{2010}]{Smithetal:10}
{Smith}, B.~D., {Hallman}, E.~J., {Shull}, J.~M.,  \& {O'Shea}, B.~W. 2010,
  ArXiv e-prints

\bibitem[\protect\citeauthoryear{{Springel}}{{Springel}}{2005}]{Springel:05}
{Springel}, V. 2005, \mnras, 364, 1105

\bibitem[\protect\citeauthoryear{{Strigari} et~al.}{{Strigari}
  et~al.}{2008}]{Strigarietal:08}
{Strigari}, L.~E., {Bullock}, J.~S., {Kaplinghat}, M., {Simon}, J.~D., {Geha},
  M., {Willman}, B.,  \& {Walker}, M.~G. 2008, \nat, 454, 1096

\bibitem[\protect\citeauthoryear{{Thoul} \& {Weinberg}}{{Thoul} \&
  {Weinberg}}{1996}]{ThoulWeinburg:96}
{Thoul}, A.~A.,  \& {Weinberg}, D.~H. 1996, \apj, 465, 608

\bibitem[\protect\citeauthoryear{{Tikhonov} \& {Klypin}}{{Tikhonov} \&
  {Klypin}}{2009}]{TikhonovKlypin:09}
{Tikhonov}, A.~V.,  \& {Klypin}, A. 2009, \mnras, 395, 1915

\bibitem[\protect\citeauthoryear{{Tinker} \& {Conroy}}{{Tinker} \&
  {Conroy}}{2009}]{TinkerConroy:09}
{Tinker}, J.~L.,  \& {Conroy}, C. 2009, \apj, 691, 633

\bibitem[\protect\citeauthoryear{{Tollerud} et~al.}{{Tollerud}
  et~al.}{2008}]{Tollerudetal:08}
{Tollerud}, E.~J., {Bullock}, J.~S., {Strigari}, L.~E.,  \& {Willman}, B. 2008,
  \apj, 688, 277

\bibitem[\protect\citeauthoryear{{Trenti} et~al.}{{Trenti}
  et~al.}{2010}]{Trentietal:10}
{Trenti}, M., {Stiavelli}, M., {Bouwens}, R.~J., {Oesch}, P., {Shull}, J.~M.,
  {Illingworth}, G.~D., {Bradley}, L.~D.,  \& {Carollo}, C.~M. 2010, \apjl,
  714, L202

\bibitem[\protect\citeauthoryear{{Tully} et~al.}{{Tully}
  et~al.}{2006}]{Tullyetal:06}
{Tully}, R., et~al. 2006, ArXiv Astrophysics e-prints

\bibitem[\protect\citeauthoryear{{Venkatesan}, {Giroux}, \&
  {Shull}}{{Venkatesan} et~al.}{2001}]{Venkatesanetal:01}
{Venkatesan}, A., {Giroux}, M.~L.,  \& {Shull}, J.~M. 2001, \apj, 563, 1

\bibitem[\protect\citeauthoryear{{Walker} et~al.}{{Walker}
  et~al.}{2009}]{Walkeretal:09}
{Walker}, M.~G., {Mateo}, M., {Olszewski}, E.~W., {Pe{\~n}arrubia}, J., {Wyn
  Evans}, N.,  \& {Gilmore}, G. 2009, \apj, 704, 1274

\bibitem[\protect\citeauthoryear{{Walker} et~al.}{{Walker}
  et~al.}{2010}]{Walkeretal:10}
{Walker}, M.~G., {McGaugh}, S.~S., {Mateo}, M., {Olszewski}, E.~W.,  \& {Kuzio
  de Naray}, R. 2010, \apjl, 717, L87

\bibitem[\protect\citeauthoryear{{Walsh}, {Jerjen}, \& {Willman}}{{Walsh}
  et~al.}{2007}]{Walshetal:07}
{Walsh}, S.~M., {Jerjen}, H.,  \& {Willman}, B. 2007, \apjl, 662, L83

\bibitem[\protect\citeauthoryear{{Walsh}, {Willman}, \& {Jerjen}}{{Walsh}
  et~al.}{2009}]{Walshetal:09}
{Walsh}, S.~M., {Willman}, B.,  \& {Jerjen}, H. 2009, \aj, 137, 450

\bibitem[\protect\citeauthoryear{{Whalen} et~al.}{{Whalen}
  et~al.}{2008}]{Whalenetal:08}
{Whalen}, D., {O'Shea}, B.~W., {Smidt}, J.,  \& {Norman}, M.~L. 2008, \apj,
  679, 925

\bibitem[\protect\citeauthoryear{{Willman} et~al.}{{Willman}
  et~al.}{2005a}]{Willmanetal:05AJ}
{Willman}, B., et~al. 2005a, \aj, 129, 2692

\bibitem[\protect\citeauthoryear{{Willman} et~al.}{{Willman}
  et~al.}{2005b}]{Willmanetal:05ApJ}
{Willman}, B., et~al. 2005b, \apjl, 626, L85

\bibitem[\protect\citeauthoryear{{Willman} et~al.}{{Willman}
  et~al.}{2010}]{Willmanetal:10}
{Willman}, B., {Geha}, M., {Strader}, J., {Strigari}, L.~E., {Simon}, J.~D.,
  {Kirby}, E.,  \& {Warres}, A. 2010, ArXiv e-prints

\bibitem[\protect\citeauthoryear{{Wise} \& {Abel}}{{Wise} \&
  {Abel}}{2007}]{WiseA:07}
{Wise}, J.~H.,  \& {Abel}, T. 2007, \apj, 671, 1559

\bibitem[\protect\citeauthoryear{{Wise} \& {Abel}}{{Wise} \&
  {Abel}}{2008}]{WiseA:08}
{Wise}, J.~H.,  \& {Abel}, T. 2008, \apj, 685, 40

\bibitem[\protect\citeauthoryear{{Wolf} et~al.}{{Wolf}
  et~al.}{2010}]{Wolfetal:10}
{Wolf}, J., {Martinez}, G.~D., {Bullock}, J.~S., {Kaplinghat}, M., {Geha}, M.,
  {Mu{\~n}oz}, R.~R., {Simon}, J.~D.,  \& {Avedo}, F.~F. 2010, \mnras, 778

\bibitem[\protect\citeauthoryear{{Zentner} et~al.}{{Zentner}
  et~al.}{2005}]{Zentneretal:05}
{Zentner}, A.~R., {Kravtsov}, A.~V., {Gnedin}, O.~Y.,  \& {Klypin}, A.~A. 2005,
  \apj, 629, 219

\bibitem[\protect\citeauthoryear{{Zucker} et~al.}{{Zucker}
  et~al.}{2006a}]{Zuckeretal:06b}
{Zucker}, D.~B., et~al. 2006a, \apjl, 650, L41

\bibitem[\protect\citeauthoryear{{Zucker} et~al.}{{Zucker}
  et~al.}{2006b}]{Zuckeretal:06a}
{Zucker}, D.~B., et~al. 2006b, \apjl, 643, L103

\end{thebibliography}

\end{document}